\documentstyle{mn}
\input{psfig}
\newcommand{\etal}{{et al.~}}

\newcommand{\kmsmpc}{\>{\rm km}\,{\rm s}^{-1}\,{\rm Mpc}^{-1}}

\newcommand{\Msun}{\>{\rm M_{\odot}}}

\newcommand{\beq}{\begin{equation}}
\newcommand{\eeq}{\end{equation}}

\newcommand{\rmd}{{\rm d}}

\newcommand{\calC}{{\cal C}}

\newcommand{\Mh}{M_{\rm h}}

\def\gtsima{$\; \buildrel > \over \sim \;$}
\def\ltsima{$\; \buildrel < \over \sim \;$}
\def\prosima{$\; \buildrel \propto \over \sim \;$}
\def\gsim{\lower.7ex\hbox{\gtsima}}
\def\lsim{\lower.7ex\hbox{\ltsima}}
\def\simgt{\lower.7ex\hbox{\gtsima}}
\def\simlt{\lower.7ex\hbox{\ltsima}}
\def\simpr{\lower.7ex\hbox{\prosima}}
\def\la{\lsim}
\def\ga{\gsim}
\def\lta{\la}
\def\gta{\ga}

\newcommand{\apj}{ApJ}

\newcommand{\aj}{AJ}
\newcommand{\mnras}{MNRAS}
\newcommand{\aap}{A\&A}

\newdimen\hssize
\newdimen\hdsize
\begin{document}


\title[Ages and Metallicities of Centrals and Satellites]
      {Ages and Metallicities of Central and Satellite Galaxies: 
       Implications for Galaxy Formation and Evolution}

\author[A. Pasquali et al.]
       {\parbox[t]{\textwidth}{
        Anna Pasquali$^{1}$\thanks{E-mail:pasquali@mpia.de},
	Anna Gallazzi$^{1}$,
        Fabio Fontanot$^{2,1}$
        Frank C. van den Bosch$^{3}$,
	Gabriella De Lucia$^{2}$,
        H.J. Mo$^{4}$,
        Xiaohu Yang$^{5}$}
        \vspace*{10pt} \\
  $^{1}$Max-Planck Institut f\"ur Astronomie, K\"onigstuhl 17,
        69117 Heidelberg, Germany\\
  $^{2}$INAF-Osservatorio Astronomico, Via Tiepolo 11, I-34131 Trieste, 
        Italy \\ 
  $^{3}$Department of Physics \& Astronomy, University of Utah,
        Salt Lake City, UT 84112-0830, USA\\ 
  $^{4}$Department of Astronomy, University of Massachusetts,
        Amherst, MA 01003-9305\\
  $^{5}$Key Laboratory for Research in Galaxies and Cosmology, 
        Shanghai Astronomical Observatory, the Partner Group of MPA, \\ 
        Nandan Road 80, Shanghai 200030, China}


\date{}
\pagerange{\pageref{firstpage}--\pageref{lastpage}}
\pubyear{2009}

\maketitle

\label{firstpage}


\begin{abstract}
  Using the stellar ages and metallicities of galaxies in the Sloan
  Digital Sky Survey (SDSS) obtained by Gallazzi et al. (2005) and the
  SDSS galaxy group catalogue of Yang et al. (2007), we study how the
  stellar ages and metallicities of central and satellite galaxies
  depend on stellar mass, $M_{\ast}$, and halo mass, $M_{\rm h}$.  We
  find that satellites are older and metal-richer than centrals of the
  same stellar mass, and this difference increases with decreasing
  $M_{\ast}$.  In addition, the slopes of the age-stellar mass and
  metallicity-stellar mass relations are found to become shallower in
  denser environments (more massive halos). This is due to the fact
  that the average age and metallicity of low mass satellite galaxies
  ($M_\ast \lta 10^{10} h^{-2} \Msun$) increase with the mass of the
  halo in which they reside.  In order to gain understanding of the
  physical origin of these trends, we compare our results with the
  semi-analytical model of Wang et al. (2008). The model, which
  predicts stellar mass functions and two-point correlation functions
  in good overall agreement with observations, also reproduces the
  fact that satellites are older than centrals of the same stellar
  mass and that the age difference increases with the halo mass of the
  satellite. This is a consequence of the fact that satellites are
  stripped of their hot gas reservoir shortly after they are accreted
  by their host halos (strangulation). The ensuing quenching of star
  formation leaves the stellar populations of satellites to evolve
  passively, while the prolonged star formation activity of centrals
  keeps their average ages younger. The resulting age offset is larger
  in more massive environments because their satellites were accreted
  earlier.  The model does not reproduce the halo mass dependence of
  the metallicities of low mass satellites, yields metallicity-stellar
  mass and age-stellar mass relations that are too shallow, and
  predicts that satellite galaxies have the same metallicities as
  centrals of the same stellar mass, in disagreement with the data.
  We argue that these discrepancies are likely to indicate the 
  need to (i)
  modify the recipes of both supernova feedback and AGN
  feedback, (ii) use a more realistic description of strangulation,
  and (iii) include a proper treatment of the tidal stripping, heating
  and destruction of satellite galaxies.
\end{abstract}




\section{Introduction}
\label{sec:intro}

Galaxies in the nearby Universe obey a number of scaling relations
between stellar mass or luminosity and rotation velocity, velocity
dispersion, size, metallicity, and black hole mass (e.g., Tully \&
Fisher 1977; Faber \& Jackson 1976; Shen \etal 2003; Tremonti \etal
2004; Magorrian \etal 1998).  At the same time, some of their
properties appear to depend on the environment in which they
reside. For instance, it is well-known that early-type galaxies are
preferentially found in dense environments, while field galaxies are
typically late-types (Oemler 1974; Dressler 1980, Postman \& Geller
1984, Whitmore, Gilmore \& Jones 1993; Goto \etal 2003; Weinmann \etal
2006a, 2009).  In addition to this morphology-density relation,
various other galaxy parameters have been shown to correlate with
their environment as well.  In particular, galaxies in denser
environment (more massive host halos) have been shown to be redder
(Hogg \etal 2004; Balogh \etal 2004b; Weinmann \etal 2006a), to have
lower star formation rates (Balogh \etal 1997, 1999; Poggianti \etal
1999; Hashimoto \etal 1998; Dominguez \etal 2002; Lewis \etal 2002;
Gomez \etal 2003; Balogh \etal 2004a; Tanaka \etal 2004; Kauffmann
\etal 2004), less nuclear activity (Miller \etal 2003; Kauffmann \etal
2004; Pasquali \etal 2009), smaller radii (Blanton \etal 2005b;
Weinmann \etal 2009), and smaller gas mass fractions (Giovanelli \&
Haynes 1985; Solanes \etal 2001; Levy \etal 2007) than those in less
dense environments.

On the other hand, it has also become clear that these same galaxy
properties correlate strongly with stellar mass and/or luminosity
(e.g., McGaugh \& de Blok 1997; Blanton \etal 2003; Baldry \etal 2004;
Kauffmann \etal 2003a,b; Hogg \etal 2004; Kelm \etal 2005; Weinmann
\etal 2006a; van den Bosch \etal 2008b; Pasquali \etal 2009).  Based
on all these findings, a picture emerges in which galaxy formation and
evolution is governed by both the stellar mass (the ``nature''
parameter) and the density or mass of the host environment (the
``nurture'' ingredient). The challenge is to disentangle the nature
effects from those induced by nurture, and to determine the
independent strength of these two parameters. This is complicated by
the fact that there exists a strong correlation between stellar mass
(luminosity) and environment, in that more massive (brighter) galaxies
preferentially reside in denser environments (Hogg \etal 2003; Mo
\etal 2004; Blanton \etal 2005b; Croton \etal 2005; Hoyle \etal 2005)
Hence, any causal connection between a galaxy property $P$ and its
stellar mass will automatically induce a correlation between $P$ and
environment, and vice versa.

One of the most common ways of defining environment is through the
projected number density of galaxies (e.g. above a given magnitude
limit).  Typically this number density, indicated by $\Sigma_n$, is
measured using the projected distance to the $n$th nearest neighbor,
with $n$ typically in the range 5-10 (e.g., Dressler 1980, Lewis \etal
2002; G\'omez \etal 2003; Goto \etal 2003; Tanaka \etal 2004; Balogh
\etal 2004a,b; Kelm \etal 2005; Cooper \etal 2008, 2009).  As
discussed in length by Weinmann \etal (2006a), the problem with this
environment indicator (or with one that uses a fixed metric aperture)
is that its physical interpretation depends on the environment itself;
in clusters, where the number of galaxies is larger than $n$, it is
representatitive only of the local environment, on a scale much
smaller than the cluster.  On the contrary, in low density
environments $\Sigma_n$ measures the global density over a spatial
scale that is much larger than the halo in which the galaxy resides.
An alternative measure of environment is provided by the
cross-correlation length of galaxies, that estimates their clustering
amplitude on scales from a few kpc to a few Mpc (cf. Wake \etal 2004;
Croom \etal 2005; Li \etal 2006; Skibba \etal 2008).  Both $\Sigma_n$
and the cross-correlation method have the disadvantage of not being
directly comparable with the environment in models of galaxy formation
and evolution, where it is usually described in terms of the dark
matter distribution, and where it is important to discriminate between
central galaxies and satellites (which are subjected to different
physical processes).  In order to allow for a more physically
intuitive description of environment, which is more directly
comparable with galaxy formation models, Yang \etal (2005) developed a
halo-based galaxy group finding algorithm which basically partitions
galaxies over dark matter halos, assigns masses to the halos, and
splits the galaxy population in centrals and satellites.

Being able to split the galaxy population in centrals and satellites
has proven to be extremely useful for investigating the impact of
satellite specific transformation processes (e.g., ram-pressure
stripping, tidal stripping, strangulation, harassment). Using the SDSS
DR4 galaxy group catalogues of Yang \etal (2007), van den Bosch \etal
(2008a) showed that, on average, satellites are redder and more
concentrated than centrals of the same stellar mass.  Under the
hypothesis that the latter are the progenitors of the former, this
suggests that some satellite specific transformation processes are at
work that make galaxies become redder and more concentrated. In fact,
van den Bosch \etal (2008a) find that central-satellite pairs matched
in both stellar mass and colour show no average concentration
difference, indicating that the transformation mechanisms affect
colour more than morphology. In addition, the colour and concentration
differences of central-satellite pairs were shown to be independent of
the mass of the halo in which the satellite resides, which implies
that satellite-specific transformation mechanisms are equally
efficient in halos of all masses. This suggests that strangulation is
most likely the main mechanism for quenching star formation in
satellites (see also van den Bosch \etal 2008b).  Weinmann \etal
(2009) took this approach a step further and analysed the radial
colour gradients of late-type satellites and centrals at fixed stellar
mass.  Satellite galaxies turn out to be smaller, fainter and redder
than centrals at nearly all galactocentric radii.  This is consistent
with a simple model, in which star formation is quenched over
time-scales of about 2 - 3 Gyr, after which the satellite galaxy is
left to evolve passively. Along these lines, Pasquali \etal (2009)
used the Baldwin, Philips \& Terlevich's (1981) diagram to separate
centrals and satellites of different activity (star formation and
optical AGN emission). They found that both star formation and AGN
activity are suppressed in satellite galaxies relative to central
galaxies of the same stellar mass, and that the dependence of
satellite `activity' (star formation or AGN activity) on halo mass is
more than four times weaker than the dependence on stellar mass.

An important shortcoming of the studies of van den Bosch \etal
(200ba,b) and Weinmann \etal (2009) is that they mainly focussed on
(broad-band) colours.  It is well-known, though, that these depend on
stellar age, metallicity and dust attenuation. Keeping dust extinction
aside, red colours may be due to an older (luminosity-weighted) age
and/or to a higher stellar metallicity. Hence, without further
information an interpretation of the colour trends mentioned above in
terms of stellar population ages and metallicities is highly
degenerate. As shown by Gallazzi \etal (2005) for the general galaxy
population in the nearby Universe, more massive galaxies are, on
average, both older and more metal rich (see also Jimenez \etal 2007).
Using a sample of early-type Bright Cluster Galaxies (BCGs, with
$M_{\rm r}$ brighter than -22 mag and $M_{\ast} >$ 10$^{11}$
M$_{\odot}$) extracted from the Sloan Digital Sky Survey (SDSS) DR6,
Bernardi (2009) showed that in any given environment satellite BCGs
are younger than central BCGs by $\sim$0.5 - 1 Gyr, while at fixed
stellar mass central and satellite BCGs are coeval. As for (stellar
and/or gas-phase) metallicity, it has been observed that, at fixed
stellar mass, galaxies in denser environment are metal-richer by only
$\sim 0.05$~dex with respect to their counterparts in low density
environments; such a small difference indicates that there is little
correlation between metallicity and environment (Sheth \etal 2006;
Mouhcine \etal 2007; Cooper \etal 2008, 2009; Ellison \etal 2009;
Loubser \etal 2009).  

In this paper we take an important next step in our study of galaxy
evolution as a function of environment: we abandon integrated galaxy
colours in favor of the actual stellar ages and metallicities derived
by Gallazzi \etal (2005).  Since the galaxy group catalogue of Yang
\etal (2007) spans $\sim$2 orders of magnitude in galaxy stellar
mass, and $\sim$4 orders of magnitude in halo mass, this allows us to
extend the above-cited studies over a much wider range of
environments.  Another important improvement of this study is that we
use semi-analytical models to interpret our findings. This paper is
organized as follows: in \S\ref{sec:data} we describe the galaxy group
catalogues and the method used to determine the ages and metallicities
of the SDSS galaxies. In \S\ref{sec:sample} we describe the general
properties of our sample of central and satellite galaxies.
\S\ref{sec:agemetal} presents the average ages and metallicities of
centrals and satellites as functions of both stellar mass and halo
mass. These results are compared to semi-analytical models in
\S\ref{sec:sam} in order to gain insight into the physical
interpretation in the context of a $\Lambda$CDM model of galaxy
formation. \S\ref{sec:disc} presents a detailed discussion of our
findings, which are summarized in \S\ref{sec:concl}.  Throughout this
paper we adopt a flat $\Lambda$CDM cosmology with $\Omega_{\rm m} =
0.238$ and $\Omega_{\Lambda} = 0.762$ (Spergel \etal 2007) and we
express units that depend on the Hubble constant in terms of $h =
H_0/(100 \kmsmpc)$.

\section{Data}
\label{sec:data}

\subsection{Galaxy groups}
\label{sec:galgroups}

The sample of galaxies analyzed in this paper is taken from the SDSS
DR4 galaxy group catalogue of Yang \etal (2007; hereafter Y07). This
group catalogue is constructed by applying the halo-based group finder
of Yang \etal (2005) to the New York University Value-Added Galaxy
Catalogue (NYU-VAGC; see Blanton \etal 2005a), which is based on SDSS
DR4 (Adelman-McCarthy \etal 2006).  From this catalogue Y07 selected
all galaxies in the Main Galaxy Sample with an extinction corrected
apparent magnitude brighter than $r=18$, with redshifts in the range
$0.01 \leq z \leq 0.20$ and with a redshift completeness $\calC_z >
0.7$.  This sample of galaxies is used to construct three group
samples: sample I, which only uses the 362,356 galaxies with measured
redshifts from the SDSS, sample II which also includes 7,091 galaxies
with SDSS photometry but with redshifts taken from alternative
surveys, and sample III which includes an additional 38,672 galaxies
that lack a redshift due to fibre-collisions, but which we assign the
redshift of its nearest neighbor (cf.  Zehavi \etal 2002).  The
analysis presented in this paper is based on sample II.  Galaxies are
split into ``centrals'', which are defined as the most massive group
members in terms of their stellar mass, and ``satellites'', which are
those group members that are not centrals.

Magnitudes and colours of all galaxies are based on the standard SDSS
Petrosian technique (Petrosian 1976; Strauss \etal 2002). They have
been corrected for galactic extinction (Schlegel, Finkbeiner \& Davis
1998), and also $K$-corrected and evolution corrected to $z=0.1$,
using the method described in Blanton \etal (2003).  We use the
notation $^{0.1}M_X$ to indicate the resulting absolute magnitude in
the photometric $X$-band.  Stellar masses for all galaxies (indicated
by $M_{\ast}$) have been computed using the relations between stellar
mass-to-light ratio and colour of Bell \etal (2003; see Y07 for
details). Galaxy structure is parameterized with the concentration
parameter $C = r_{90}/r_{50}$, where $r_{90}$ and $r_{50}$ are the
radii that contain 90$\%$ and 50$\%$ of the Petrosian $r$-band flux,
respectively. As shown by Strateva \etal  (2001), $C$ is a reasonable
proxy for Hubble type, with $C >$ 2.6 selecting mostly bulge-dominated
galaxies. 

For each group in the Y07 catalogue two estimates of its dark matter
halo mass, $\Mh$, are available: one based on the ranking of its total
characteristic luminosity, and the other based on the ranking of its
total characteristic stellar mass. Both halo masses agree very well
with each other, with an average scatter that decreases from $\sim
0.1$~dex at the low mass end to $\sim 0.05$~dex at the massive
end. With the method of Y07, halo masses can only be assigned to
groups more massive than $\sim 10^{12} h^{-1} \Msun$ which have at
least one member with $^{0.1}M_r - 5 \log h \leq$ -19.5 mag. For
smaller mass halos, Yang \etal (2008) have used the relations
between the luminosity (stellar mass) of central galaxies and the halo
mass of their groups to extrapolate the halo mass of single central
galaxies down to $\Mh \simeq 10^{11} h^{-1} \Msun$. This extends the
number of galaxies with an assigned halo mass from 295,861 in the
original Y07 paper to all 369,447 galaxies in sample II. In what
follows, we will make use of the halo masses obtained from the group's
characteristic stellar mass\footnote{We have verified, though, that
  none of our results change significantly if we adopt the
  luminosity-rank based masses instead.}.

This sample is not volume-limited, and thus suffers from Malmquist
bias, causing an artificial increase of the average luminosity (and
also stellar mass) of galaxies with increasing redshift. This effect
is more severe for satellites which, within each halo, have a wide
mass distribution.  To correct for this bias, we weight each galaxy by
$1/V_{\rm max}$, where $V_{\rm max}$ is the comoving volume of the
Universe out to a comoving distance at which the galaxy would still
have made the selection criteria of our sample.  In what follows all
distributions are weighted by $1/V_{\rm max}$, unless specifically
stated otherwise. Note, though, that since we will always present
our results for narrow bins in stellar mass, none of our results are
sensitive to this particular weighting scheme; in fact, none of our
results change qualitatively if we use no weighting at all.
\begin{figure}
\centerline{\psfig{figure=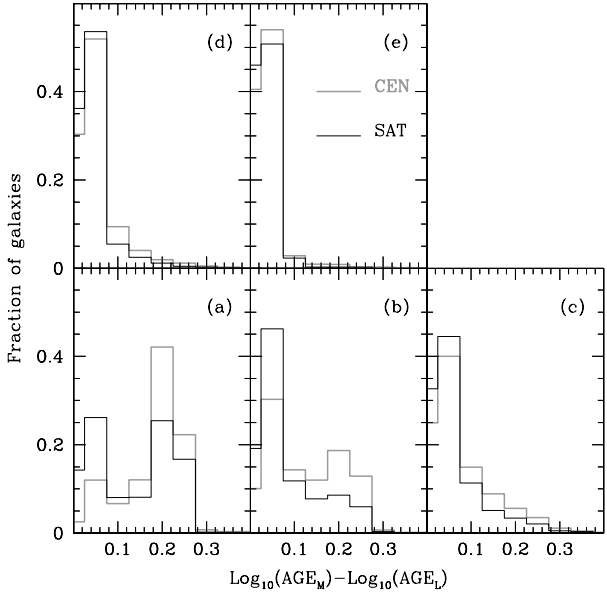,width=\hssize}}
\caption{Normalized distributions of the difference between
  mass-weighted and luminosity-weighted ages of central galaxies
  (grey) and satellite galaxies (black). Results are shown for
  different logarithmic bins in stellar mass (in $h^{-2}\Msun$); 
  panel (a): 9--9.5, panel (b): 9.5--10, panel (c): 10--10.5, 
  panel (d): 10.5--11, panel (e): $>$11).}
\label{fig:agem_agel}
\end{figure}

\subsection{Galaxy stellar populations}
\label{sec:galstelpop}

We have matched sample II with the catalogue of stellar ages and
metallicities of SDSS galaxies by Gallazzi \etal (2005). For each
galaxy, Gallazzi \etal  (2005) determined the full probability
density function (PDF) of stellar $r$-band flux-weighted age and
metallicity: the median of the PDF represents the fiducial estimate of
the parameter, while the associated uncertainty is given by half of
the 16th-84th percentile range of the PDF. The PDF of each parameter
has been derived by comparing the observed strength of spectral
absorption features with the predictions of a Monte Carlo library of
150000 star formation histories (SFHs), based on the Bruzual \&Charlot
(2003) population synthesis code and the Chabrier (2003) Initial Mass
Function (IMF).  The SFHs in the library are modelled by an
exponentially declining star formation rate (SFR), with varying time
of onset and timescale, to which random bursts of varying intensity
and duration are superposed with a probability that allows 10\% of
models to experience a burst in the last 2~Gyr. The set of absorption
features used to constrain the PDF includes the 4000\AA-break and the
Balmer lines as age-sensitive indices, and $\rm [Mg_2Fe]$ and $\rm
[MgFe]^\prime$ as metal-sensitive indices.

The derived stellar ages and metallicity refer to the redshift
at which the galaxies were observed. Given the (small) redshift range
of the sample and the magnitude selection of the survey, this could
potentially bias the relations as a function of stellar mass. However,
we do not attempt to correct for this effect because it
would require an accurate knowledge of
the star formation history from the redshift of the observations to the
present. Under the simplest assumption of passive evolution (which does not
apply to the entire sample), the stellar ages could be all scaled
to $z =$ 0 by adding the look-back time. Such a correction would steepen
the age - stellar mass relation.

The uncertainty on the derived stellar metallicity depends strongly on
the spectral signal-to-noise (S/N). Specifically, a median S/N per
pixel of at least 20 is required to constrain stellar metallicity
within $0.3$~dex.  Furthermore, at fixed S/N, the broadness of the
metallicity PDF varies as a function of galaxy type, being larger for
low mass, star-forming galaxies: these galaxies generally have lower
metallicities and hence more difficult to measure Fe and Mg absorption
features. The quality of the spectrum is a less stringent requirement
for light-weighted age estimates, which have a typical uncertainty of
$0.12$~dex with little dependence on galaxy type.

Note that the SDSS spectra have been aquired with a 3~arcsec-diameter
fibre and thus sample preferentially the inner $\sim 60\%$ of the
galaxy light distribution. This can affect estimates of global stellar
age and metallicity, depending on the strength of stellar population
gradients in galaxies. Gallazzi et al. (2005) did not detect any significant 
bias in age as a function of normalized redshift at fixed stellar mass, except 
for the most massive late-type galaxies. Similarly, they estimated that 
the measured metallicity of intermediate-mass bulge dominated galaxies would 
vary by $\la 0.2$~dex if such a galaxy
would be moved from one edge of the survey to the other. While we caution that 
individual stellar age and metallicity estimates may be biased because of 
stellar populations gradients, we have checked that the mean age/metallicity -
 mass relations discussed in this work are not affected by variations in the
fibre covering factor (see Sect. 4). 

Another source of systematic uncertainty is the variation in element
abundance ratios. Although the stellar parameter estimates are based
on absorption indices which are to first order insensitive to
variations in $\alpha$/Fe, the stellar metallicity could be
overestimated by no more than $\sim 0.05$~dex. For massive early-type
galaxies, which are known to be $\alpha$-enhanced, the light-weighted
ages can be underestimated by a similar amount (Gallazzi \etal
2005). Finally, the prior according to which starbursts are generated
in the Monte Carlo library can also have an effect on the derived
parameters. For example, by increasing the probability of having
undergone a burst in the last 2~Gyr to 50\% would result in ages that
are younger by $\sim 0.07$~dex and metallicities that are higher by
$\sim 0.04$~dex. Note, however, that this would mainly affect
early-type galaxies dominated by old stellar populations. For more
details and a more elaborate discussion of systematic effects we refer
the reader to Gallazzi \etal (2005) and Gallazzi \etal (2008).
\begin{figure*}
\centerline{\psfig{figure=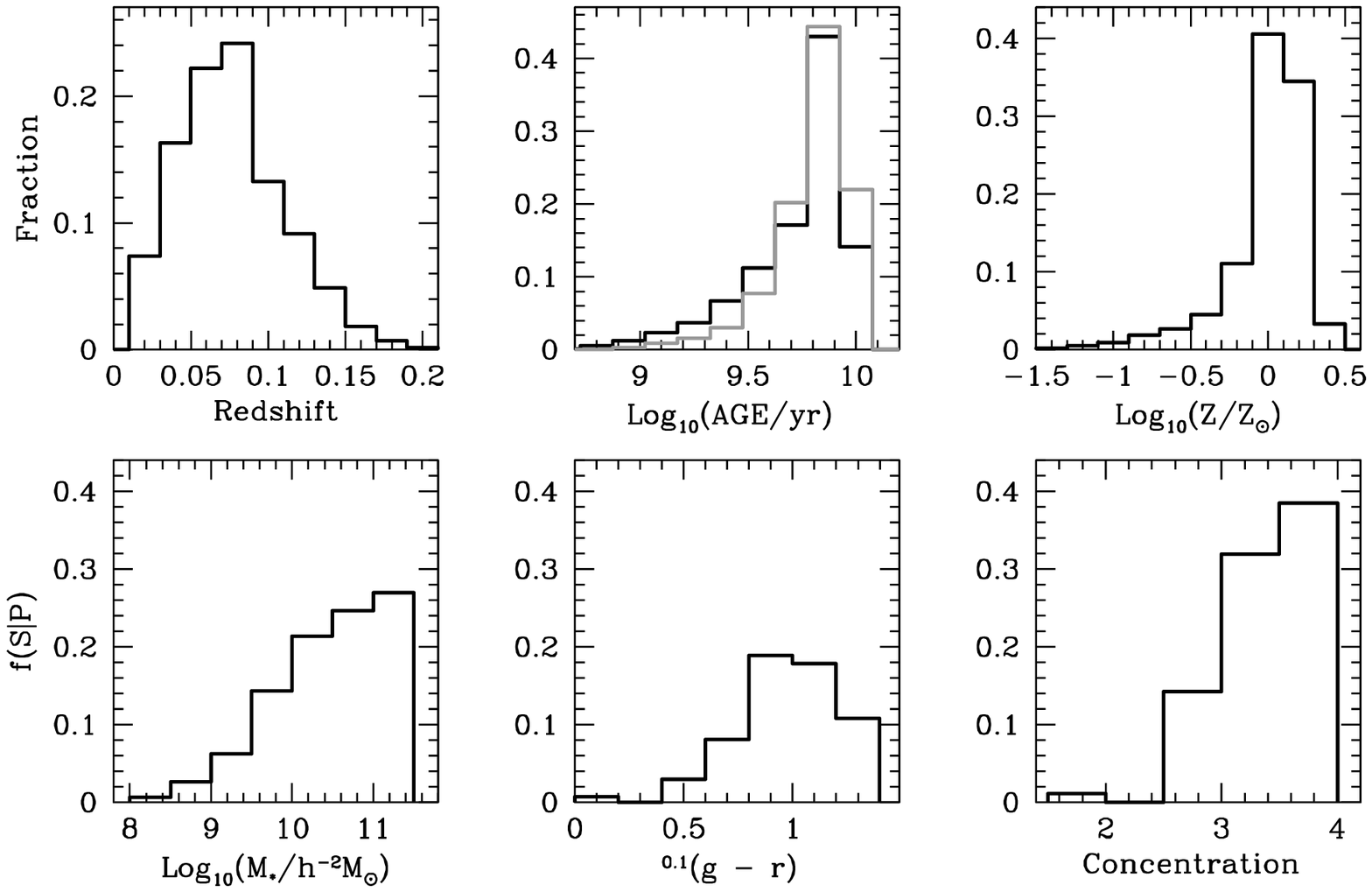,width=0.8\hdsize}}
\caption{The upper panels show the distribution of galaxies in sample
  $S$ in redshift (left-hand panel), stellar age (middle panel;
  luminosity- and mass-weighted in black and grey, respectively) and
  metallicity (right-hand panel). The bottom panels show the weighted
  fractions $f(S|P)$ of galaxies in sample II that are part of sample
  $S$ as function of three galaxy properties $P$: stellar mass
  ($M_{\ast}$; left-hand panel) colour ($^{0.1}(g - r)$, middle panel)
  and concentration ($C$, right-hand panel).}
\label{fig:sample}
\end{figure*}

In this work we complement the luminosity-weighted ages with estimates
of the mass-weighted age. For each model in the SFH library the
mass-weighted age is computed by weighting each generation of stars by
their mass, taking into account the fraction of mass returned to the
interstellar medium by long-lived stars. The mass-weighted ages of
observed galaxies is then estimated in the same way as the other
parameters as described above (for more details see Gallazzi \etal
2008 where this quantity has been derived and used). Uncertainties on
the two age estimates are comparable. While the luminosity-weighted
age is more sensitive to small fractions of recent generations of
stars (which contribute significantly in luminosity but not in mass),
the mass-weighted age is more representative of the average epoch when
the bulk of the stars in a galaxy formed. The mass-weighted age is
always older than the luminosity-weighted age and their difference can
give (at least qualitative) insight into the recent star formation
history. 

Fig.~\ref{fig:agem_agel} shows the $1/V_{\rm max}$-weighted histograms
of the difference between the mass-weighted and luminosity-weighted
ages for centrals (grey) and satellites (black), in different bins of
galaxy stellar mass.  At high stellar masses (roughly above
Log$_{10}(M_{\ast}/h^{-2}$M$_{\odot})=$10.5) the mass-weighted ages of
both centrals and satellites are typically only $\sim 0.05$~dex older
than their luminosity-weighted ages. At lower stellar masses, however,
there is a more pronounced tail toward larger differences between
mass-weighted and luminosity-weighted ages. A second peak at an
age-difference of $\sim0.2$~dex ($\sim 1.6$~Gyr) is clearly visible in
galaxies with $M_\ast\lta 10^{10} h^{-2}$M$_\odot$. This peak is much
more populated in the central galaxy population. This hints at a
typically more prolongued SFH in low mass central galaxies with
respect to satellite galaxies of the same stellar mass. We will return
to this in \S\ref{sec:agemetal}.

\section{Basic properties of the sample}
\label{sec:sample}

For our analysis we select galaxies in sample II with known age and
metallicity and whose spectroscopic S/N per pixel is larger than
20. Such a cut in S/N effectively selects galaxies with a
1$\sigma$ error of $<0.1$~dex on age and $<0.2$~dex on stellar
metallicity. These objects are hereafter referred to as the
spectroscopic ($S$) sample, which contains a total of 70,067 galaxies,
split between 56,441 centrals and 13,626 satellites.  As shown in the
upper panels of Fig.~\ref{fig:sample}, the galaxies in sample $S$ span
the redshift range 0.01 $\leq z \leq$ 0.20 as set by the Y07 group
catalogue. The distribution of their luminosity-weighted ages (in
black) ranges from between $\sim$600 Myr to $\sim 10$~Gyr and peaks at
$\sim$6~Gyr. The same is true for the mass-weighted ages (in grey),
although they are shifted towards somewhat larger values. The
distribution of stellar metallicities covers the interval -1.5 $\lta$
Log$_{10}(Z/Z_{\odot}) \lta$ 0.5 and peaks at Solar metallicity.

In order to assess how representative the $S$ sample is of the full
galaxy population in the Y07 groups, the bottom panels of
Fig.~\ref{fig:sample} show $f(S|P)$, defined as the fraction of
galaxies with properties $P$ (stellar mass, colour or concentration)
that are present in sample $S$:
\begin{equation}
f(S|P) = \sum\limits_{i=1}^{N_{S|P}} w_i /
         \sum\limits_{i=1}^{N_P} w_i\,.
\end{equation}
Here $w_i = 1/V_{{\rm max},i}$ is the weight of galaxy $i$, $N_{S|P}$
is the number of galaxies in sample $S$ with properties $P$ and $N_P$
is the number of galaxies in sample II with properties $P$.  This
shows that the requirement of a S/N per pixel of 20 or higher biases
the sample towards massive, red, early-type galaxies.  However, since
the main goal of this paper is to compare the ages and metallicities
of central galaxies to those of satellites, rather than study age
and/or metallicity distributions, this bias will not have a signficant
impact on our results. 

\begin{figure*}
\centerline{\psfig{figure=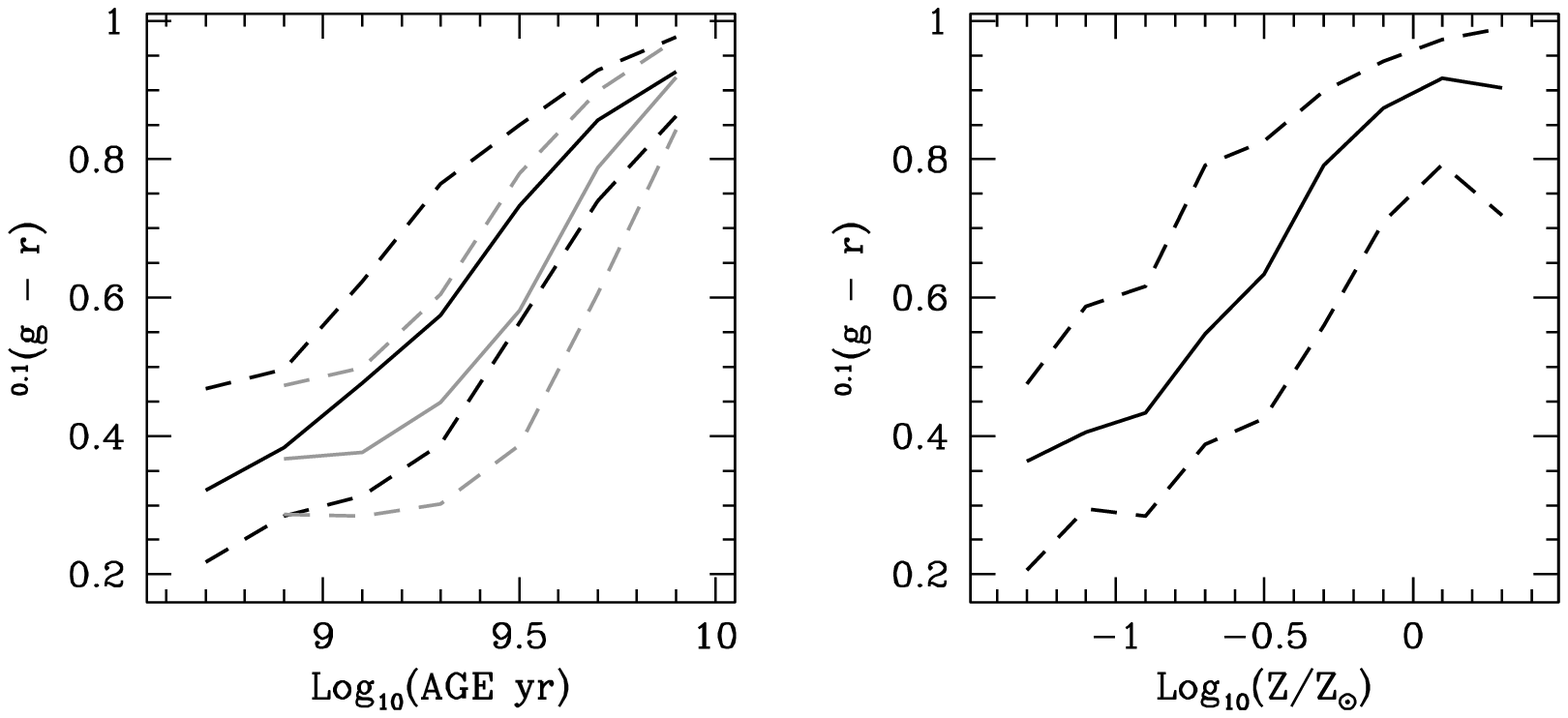,width=0.8\hdsize}}
\caption{The $^{0.1}(g - r)$ colour distribution as a function of
  stellar age (left-hand panel; luminosity- and mass-weighted in black
  and grey, respectively) and metallicity (right-hand panel) for the
  galaxies in sample $S$. The solid line represents the median colour
  (metallicity), while the dashed lines trace the 16th and 84th
  percentiles of the colour distribution. Redder galaxies are both
  older and more metal-rich, highlighting an important shortcoming of
  using colours to probe the stellar mass/environment dependence of
  the galaxy population.}
\label{fig:colours}
\end{figure*}

The solid lines in Fig.~\ref{fig:colours} show the average
$^{0.1}(g-r)$ colour of galaxies in sample $S$ as function of both
their age (left-hand panel; luminosity and mass-weighted ages in black
and grey, respectively) and metallicity (right-hand panel). The dashed
lines show the corresponding 16th and 84th percentiles of the colour
distribution.  This shows the well-know fact that redder galaxies are
both older and more metal-rich, and emphasizes that studies of the
colour dependence of galaxies as function of stellar mass and/or
environment (as for example in Hogg \etal 2004; Blanton \etal 2004;
Weinmann \etal 2006a; van den Bosch \etal 2008) cannot discriminate
between age and metallicity effects. This highlights the main
improvement of this paper with respect to aforementioned studies.

\section{Ages and Metallicities of Centrals \& Satellites}
\label{sec:agemetal}

Since the main aim of this paper is to determine whether satellite and
central galaxies differ in their stellar age and metallicity and how
these differences depend on environment, we now split the galaxies in
sample $S$ in centrals and satellites and analyse the distributions of
their stellar ages (luminosity- and mass-weighted, $AGE_{\rm L}$ and
$AGE_{\rm M}$, respectively) and metallicities as a function of their
stellar mass, $M_{\ast}$, and halo mass, $M_{\rm h}$.

\begin{figure*}
\centerline{\psfig{figure=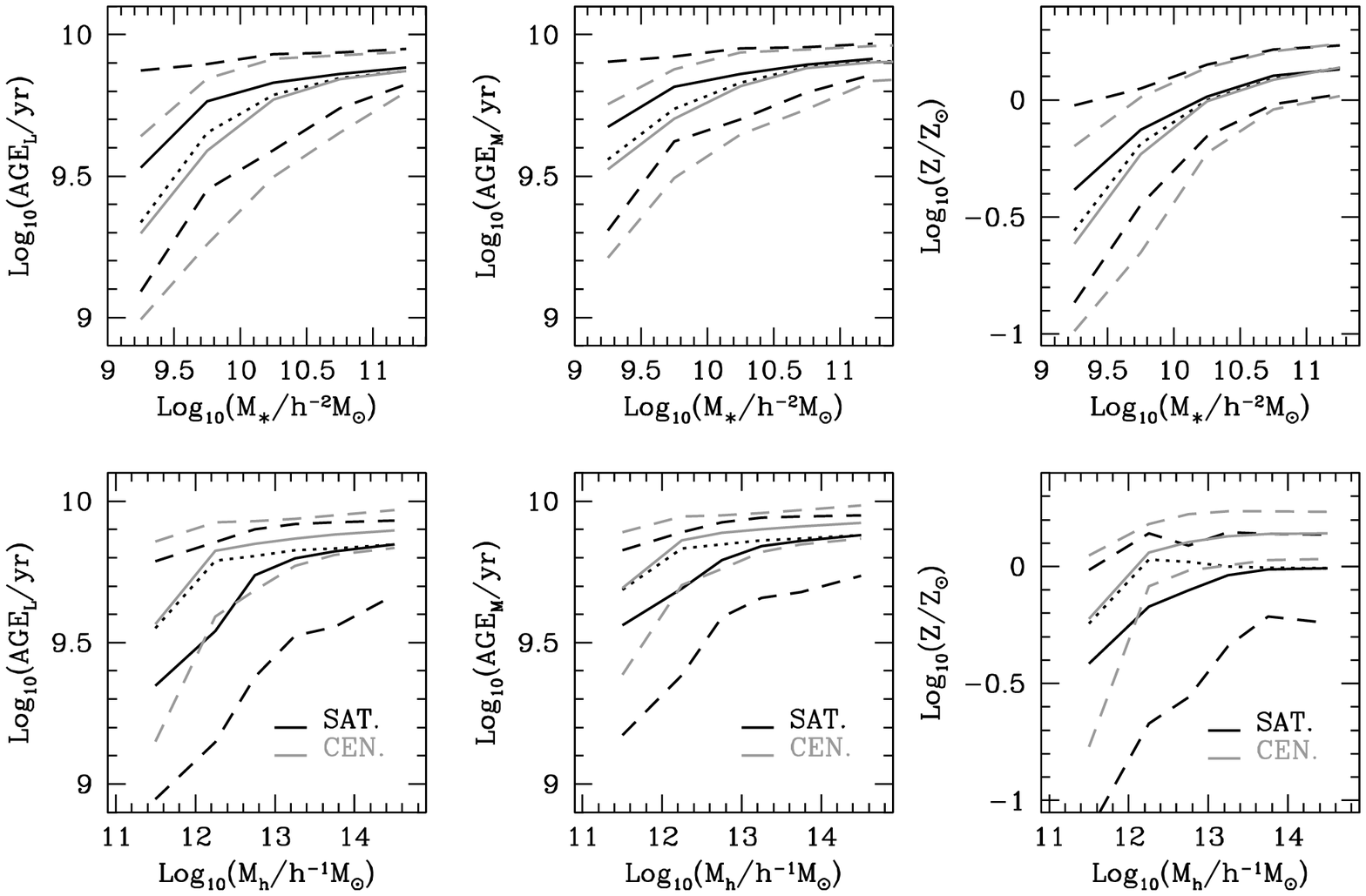,width=0.95\hdsize}}
\caption{Distribution of stellar age and metallicity as function of
  stellar mass (upper panels) and halo mass (lower panels) for central
  (grey) and satellite (black) galaxies. From left to right the panels
  show the luminosity-weighted ages, the mass-weighted ages, and the
  stellar metallicities. Solid lines correspond to the median while
  the dashed lines indicate the 16th and 84th percentiles of the
  corresponding distributions. The black, dotted line represents the 
  median age (metallicity) of all galaxies in the $S$ sample.}
\label{fig:compCS}
\end{figure*}

The top panels of Fig.~\ref{fig:compCS} show the age- and
metallicity-stellar mass relations for centrals and satellites, in
grey and black respectively. As in Fig.~\ref{fig:colours}, the solid
and dashed lines indicate the median and the 16th and 84th percentiles
of the age (metallicity) distributions, respectively. For comparison,
the black, dotted line indicates the median age (metallicity) of {\it
all} galaxies in sample $S$ (centrals and satellites combined).  As
already noted in Fig.~\ref{fig:sample}, the mass-weighted age
distributions are shifted to older ages by $\sim 0.2$~dex at
Log$_{10}(M_{\ast}/h^{-2}$M$_{\odot}$) $<$ 10.5 but coincide with the
luminosity-weighted age distributions at larger stellar masses.

To test the effects of aperture bias we have computed mean relations for
centrals and satellites in narrow bins of the ratio between the fibre radius and
the Petrosian half-light radius ($R_{\rm fibre}/R_{50,r}$). We find that for 
galaxies in which $R_{\rm fibre}/R_{50,r}<0.7$ the age/metallicity--mass relations 
are somewhat steeper. These galaxies represent 31\% and 28\% of all centrals 
and satellites in the $S$ sample respectively, and do not dominate the
global relations shown in Fig.~\ref{fig:compCS}.
For those galaxies in which the fibre samples a higher fraction 
of the light (up to $R_{\rm fibre}/R_{50,r}=2$) the relations at fixed
$R_{\rm fibre}/R_{50,r}$  agree well with the global relations shown in
Fig.~\ref{fig:compCS}. We are thus confident that the shape of the relations is
not significantly affected by aperture bias. This holds for both centrals and
satellites separately. 

From Fig..~\ref{fig:compCS} there is a clear indication that satellite galaxies have older stellar
populations than centrals of the same stellar mass, and this age
difference increases with decreasing stellar mass. At the massive end
($M_{\ast} \gta$ 10$^{11} h^{-2}$M$_{\odot}$) satellites and centrals
are equally old on average, while at the low mass end ($M_{\ast} \sim$
3 $\times$ 10$^{9} h^{-2}$M$_{\odot}$) the difference is $\sim 1$~Gyr
(compared to a 1$\sigma$ scatter of $\sim 3$~Gyr).  In absolute terms,
the age differences are very similar when using luminosity-weighted or
mass-weighted ages, suggesting that they are not merely due to
differences in the {\it recent} (i.e. past 1-2 Gyrs) star formation
history. In terms of metallicity, massive satellite galaxies have a
similar metallicity as centrals of the same stellar mass.  However, at
the low mass end, the metallicity of satellites is higher than that of
centrals (by $\sim 0.1$~dex at $M_{\ast} =$ 10$^{9}
h^{-2}$M$_{\odot}$).  

Comparing the 16th and 84th percentiles of the age distributions,
one notices (i) that the age distribution of low mass satellites has
an excess at old ages compared to the age distribution of low mass
centrals, and (ii) a relative lack of young satellites compared to
centrals at all stellar masses.  Similar features are also present in
the metallicity distributions, which show an excess of metal-rich
satellites at $M_{\ast} <$ 6 $\times$ 10$^{9} h^{-2}$M$_{\odot}$ and a
relative lack of metal-poor satellites at all stellar masses.

A comparison with the black dotted line shows that the median age
(metallicity) of all galaxies in sample $S$ closely follows that of
centrals.  This simply reflects the fact that centrals are more
numerous than satellites in any $M_{\ast}$-bin.

The bottom panels of Fig.~\ref{fig:compCS} show a similar comparison
of the age- and metallicity-distributions of centrals vs. satellites,
but now as function of halo mass rather than stellar mass. As is
evident, in any given environment (i.e. halo mass), centrals are
systematically older (by $\sim 1$~Gyr) and metal-richer (by $\sim
0.15$~dex) than satellites. The former is in excellent agreement with
Bernardi (2009).  These results, which hold for the median, the 16th
and 84th percentiles of the age (metallicity) distributions, are
most likely reflections of the age- and metallicity-$M_{\ast}$
relations, since, by definition, a central galaxy is more massive than
its satellite galaxies.

A comparison of the lower and upper panels of Fig.\ref{fig:compCS}
shows that for central galaxies the dependence of stellar age and
metallicity on halo mass is significantly weaker than that on stellar
mass. In fact, the age and metallicity distributions of centrals are
almost independent of halo mass for $M_{\rm h} >$ 10$^{12}
h^{-1}$M$_{\odot}$. In the case of satellite galaxies, the ages and
metallicities of their stellar populations seem to depend on $M_{\rm
h}$ and $M_{\ast}$ with similar strength.

The black dotted curves in the lower panels of Fig.\ref{fig:compCS}
show that the median age and metallicity of {\it all} galaxies in
sample $S$ (central and satellites combined) are almost independent of
halo mass for $M_{\rm h} >$ 10$^{12} h^{-1}$M$_{\odot}$. This is due
to the fact that centrals outnumber satellites in halos less massive
than Log$_{10}(M_{\rm h}/h^{-1}$M$_{\odot}) \sim$ 13.5, while
satellites constitute the numerical majority in more massive groups.
These results are in good agreement with Sheth \etal (2006) and
Ellison \etal (2009), who found that the gas-phase and stellar
metallicities of galaxies are virtually independent of environment.
Our results show that this lack of environment dependence is somewhat
fortuitous, and that a more pronounced dependence of stellar age and
metallicity on environment emerges when centrals and (especially)
satellites are treated separately.

The results in Fig.\ref{fig:compCS} show that the ages and
metallicities of centrals and satellites scale with both halo mass and
stellar mass.  In order to make progress, and to determine which of
these dependencies is causal, we need to investigate the halo mass
dependence at fixed stellar mass and vice versa.  Since there is
relatively little scatter in the relation between $M_{\rm h}$ and
$M_{\ast}$ for centrals, there is not sufficient dynamic range to
determine which of these two parameters causally controls the ages
and/or metallicities of central galaxies (cf. Pasquali \etal 2009).
However, in the case of satellite galaxies, the dynamic range of halo
masses occupied by satellites of a given stellar mass can be large
(especially for low mass systems), thus allowing for a detailed
causality study (see e.g. van den Bosch \etal 2008b).

\begin{figure*}
\centerline{\psfig{figure=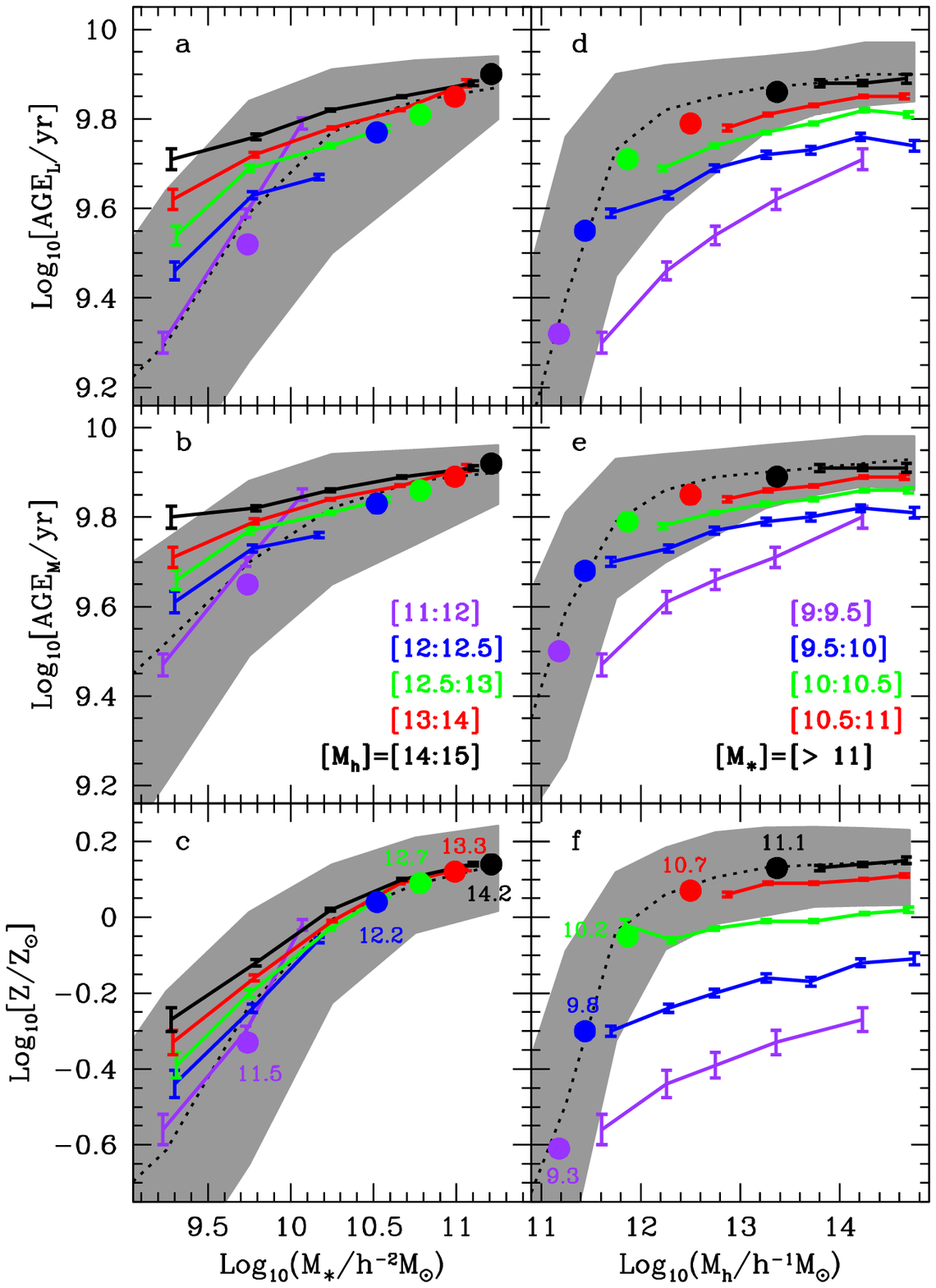,width=0.8\hdsize}}
\caption{Ages and metallicities of central and satellite galaxies as
  function of stellar mass (left-hand panels) and halo mass
  (right-hand panels). From top to bottom the panels show the
  luminosity-weighted ages (panels $a$ and $d$), the mass-weighted
  ages (panels $b$ and $e$) and the stellar metallicities (panels $c$
  and $f$).  The grey band in each panel marks the 16th-to-84th
  percentile range of the distribution of centrals, with the dotted
  line indicating the corresponding median. The coloured solid lines
  indicate the medians for satellites in logarithmic bins of halo mass
  (left-hand panels) or stellar mass (right-hand panels), as indicated
  in panels $b$ and $e$ (stellar and halo masses are in $h^{-2} \Msun$
  and $h^{-1} \Msun$, respectively). The coloured filled circles
  indicate the corresponding medians for the centrals in those bins,
  with the associated mean logarithmic values of $M_{\rm h}$ and
  $M_{\ast}$ indicated in panels $c$ and $f$, respectively.}
\label{fig:res}
\end{figure*}

The solid, coloured lines in Fig.\ref{fig:res} show the mean ages and
metallicities of satellite galaxies as function of stellar mass for
narrow bins in halo mass (left-hand panels), and as function of halo
mass for narrow bins in stellar mass (right-hand panels). The
errorbars indicate the errors on the mean, where each point contains
at least 30 satellite galaxies.  In each panel, the grey band depicts
the range of ages/metallicities enclosed by the 16th and 84th
percentiles of the corresponding distributions for centrals, with the
black dotted line indicating the corresponding median.

A comparison of panels $a$ and $d$ shows that massive satellite
galaxies ($M_{\ast} \gta 3 \times 10^{10} h^{-2} \Msun$) have
luminosity weighted ages that are (i) comparable to those of central
galaxies of the same mass, and (ii) virtually independent of the mass
of the halo in which they reside. However, in the case of low mass
satellites ($M_{\ast} \lta 3 \times 10^{10} h^{-2} \Msun$), there is a
clear dependence on halo mass, such that satellites in more massive
halos have older stellar populations. This halo-mass dependence is
most pronounced for the satellites in the lowest stellar mass bin probed
here ($9 < {\rm Log}_{10}(M_{\ast}/h^{-2}\Msun) \leq 9.5$), which have
an average luminosity weighted age that increases from $\sim 2$~Gyr in
halos with $11 < {\rm Log}_{10}(M_{\rm h}/h^{-1}\Msun) \leq 12$ (very
similar to that of central galaxies of the same stellar mass) to $\sim
5$~Gyr in massive clusters.

Panels $b$ and $e$ show similar results for the mass-weighted ages.
Once again, massive satellites have stellar ages that are comparable
to those of central galaxies of the same mass, and independent of halo
mass (environment). In the case of low mass satellites, their stellar
ages increase with the mass of the halo in which they reside; in the
lowest stellar mass bin probed, we again find a difference of $\sim
3$~Gyr between the (mass-weighted) ages of satellites in halos with
$11 < {\rm Log}_{10}(M_{\rm h}/h^{-1}\Msun) \leq 12$ and those in
massive clusters.

Analogous trends can be seen for the stellar metallicities, shown in
panels $c$ and $f$. Massive satellites with $M_{\ast} \gta 3 \times
10^{10} h^{-2} \Msun$, which all reside in halos more massive than
$\sim 3\times 10^{12} h^{-1}\Msun$, have metallicities that are
similar to centrals of the same stellar mass and independent of halo
mass. The metallicities of satellites with $M_{\ast} \lta 3 \times
10^{10} h^{-2} \Msun$, though, increase with the mass of the halo in
which they reside. Similar as for the ages, this mass dependence becomes
stronger for less massive satellites.  In the case of satellites with
$9 < {\rm Log}_{10}(M_{\ast}/h^{-2}\Msun) \leq 9.5$, the mean stellar
metallicity increases by almost 0.3~dex from ${\rm
  Log}_{10}[Z/Z_{\odot}] \simeq -0.55$ in halos with $11 < {\rm
  Log}_{10}(M_{\rm h}/h^{-1}\Msun) \leq 12$ (very similar to that of
central galaxies of the same stellar mass) to ${\rm
  Log}_{10}[Z/Z_{\odot}] \simeq -0.27$ in massive clusters.
\begin{figure*}
\centerline{\psfig{figure=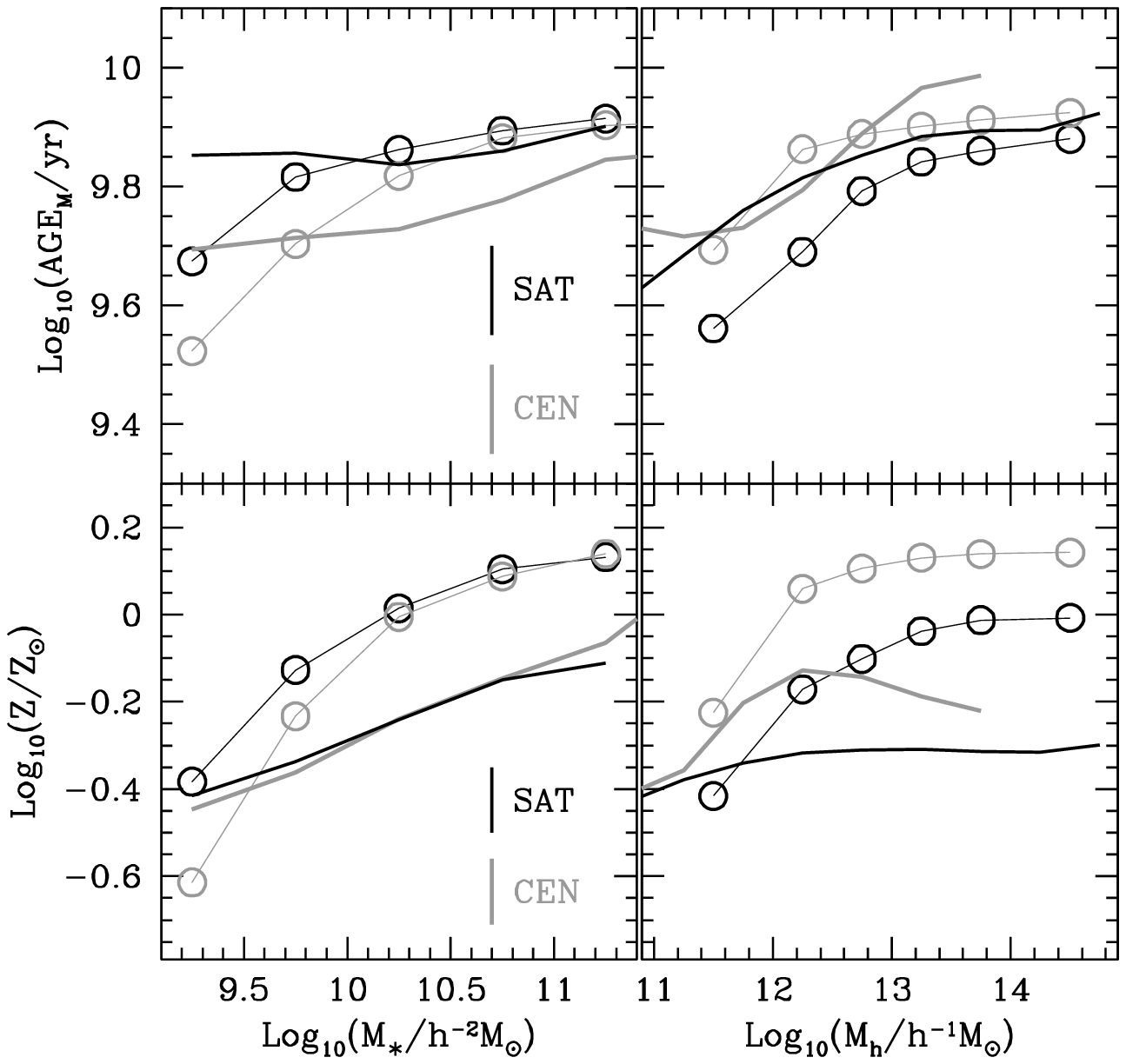,width=0.7\hdsize}}
\caption{The average mass-weighted age (upper panels) and metallicity
  (lower panels) of central (grey solid lines) and satellite (black
  solid lines) galaxies as functions of stellar mass (left-hand
  panels) and halo mass (right-hand panels) in the SAM of Wang \etal
  (2008). The typical $\pm 1\sigma$ scatter is indicated by the
  vertical bars in the left-hand panels, while the typical error on
  the mean is smaller than a circle. For comparison, the grey and
  black open circles indicate the observed values of centrals and
  satellites, respectively, in our sample $S$
  (cf. Fig.~\ref{fig:compCS}).} 
\label{fig:modelCS}
\end{figure*}

We have tested the robustness of the results obtained so far by
performing the same analysis on galaxy subsamples that are volume
limited and complete in stellar mass (according to eq.~[A8] in van den
Bosch \etal 2008a). These tests confirm the trends seen in
Fig.\ref{fig:res}.  Also, the results of Figs.~\ref{fig:compCS}
and~\ref{fig:res} do not change if (i) we use all galaxies for which
ages and metallicities are available, irrespective of whether their
spectra have a S/N per pixel larger than 20, or (ii) we use velocity
dispersion (corrected for aperture) rather than stellar mass.

\section{Comparison with semi-analytical model}
\label{sec:sam}

The main results from the previous section are: 
\begin{itemize}
\item At fixed stellar mass, satellites are older and metal-richer
  than centrals, and the difference increases with decreasing
  $M_{\ast}$.
\item The ages and metallicities of massive satellites are independent
  of their environment, while those of low mass satellites increase
  with increasing halo mass.
\end{itemize}
Clearly, these findings hold important clues to how galaxies form and
evolve in different environments. In this section we compare the
observational results with predictions of a semi-analytical model.
This comparison serves two purposes. First of all, we want to
investigate how well a state-of-the-art semi-analytical model, which
fits many of the {\it global} properties of the galaxy population
(stellar mass function, colour-magnitude relation, clustering
properties), can reproduce the age- and metallicity-trends revealed
here. Secondly, we want to use the model to gain insight into the
physical processes that underlie these trends.

\subsection{Model description}
\label{sec:modeldescr}

Semi-analytical models (hereafter SAMs) anchor galaxy formation to the
large scale matter distribution as traced by dark matter halos and
follows the evolution of their baryonic content using simplified, yet
physically and/or observationally motivated prescriptions for the
treatment of gas cooling, star formation, supernovae and AGN feedback,
and galaxy mergers (see Baugh 2006 for a comprehensive review). In
recent years, these models have become a widely used tool to predict
statistical properties of galaxies to be compared with data from
modern multi-wavelength surveys. Despite much encouraging success, a
number of discrepancies still exist between SAM predictions and
observational results. For example, it remains challenging to fit the
faint-end slope of the galaxy luminosity function (e.g. Benson \etal
2003; Mo \etal 2005), and the models typically predict disk rotation
velocities that are too high, unless adiabatic contraction and/or disk
self-gravity are ignored (e.g. Cole \etal 2000; Dutton \etal 2007). In
addition, SAMs have problems matching the evolution of the galaxy mass
function with redshift (e.g.,De Lucia \& Blaizot 2007; Somerville
\etal 2008; Fontanot \etal 2009). Most relevant for this
discussion, SAMs typically overpredict the red fraction of satellite
galaxies (Baldry \etal 2006; Weinmann \etal 2006b; Kimm \etal 2009)
as well as the stellar mass density in low-mass galaxies (Gallazzi et
al. 2008). Comparing the observed distribution in stellar mass density
as a function of age/metallicity with the one predicted by the Millenium 
Simulation, Gallazzi et al. (2008) found a clear deficit of young galaxies 
and a too narrow metallicity range in the simulation.

In this paper, we use results from the ``Munich'' semi-analytic model
described in Wang \etal (2008, hereafter W08). This is essentially
the same model as that discussed in De Lucia \& Blaizot (2007), but
adapted to a cosmology in better agreement with the third-year data
release of the WMAP mission (Spergel \etal 2007). This model has been
used extensively in a number of recent studies and has been shown to
provide nice agreement with various observational measurements, both
in the local Universe and at higher redshift. For more details on the
physical modelling of the various processes considered, we refer to
the original paper and references therein. Here, we briefly summarize
those modelling details that are relevant for this study.

The model we use follows dark matter substructures, which allows us to
follow the motion of the galaxies sitting at their centers until tidal
truncation and stripping disrupt the subhalos at the resolution limit
of the simulation. When this happens, a ``residual merging time'' is
estimated from the current orbit and the classical dynamical friction
formula (see De Lucia \& Blaizot 2007 for details). The positions and
velocities of ``orphan'' galaxies (those that are no longer associated
with a distinct subhalo) are estimated following the most bound
particle of the parent dark matter substructure at the last time it
was identified. The galaxy associated with a main halo is defined as
{\it central}, while all others are labelled {\it satellites}
regardless of whether they are still associated with a subhalo (in the
simulation) or they are orphans.  When a central galaxy becomes a
satellite, its reservoir of hot gas is assumed to be instantaneously
stripped and added to the hot component associated with the central
galaxy. It has already been noted that this produces a rapid decline
of the star formation activity and reddening of the stellar population
(Weinmann \etal 2006b; see also Fontanot \etal 2009). Like most of
the semi-analytical models published to date, this model adopts the
instantaneous recycling approximation for chemical evolution. Upon
their synthesis, metals are instantaneously returned to the cold gas
with a 100\% mixing efficiency. Finally, the model also tracks the
formation of supermassive black holes. It differentiates between a
merger-induced ``quasar-mode'', during which the black hole (the
merged product of the black holes in the progenitor galaxies) accretes
a certain fraction of the cold gas present in the progenitor galaxies,
and the ``radio-mode'', during which the black hole accretes hot gas.
It is assumed that this ``radio-mode'' accretion results in energy
feedback into the surrounding medium which reduces or stops the
cooling flow (AGN feedback). Note, though, that the model does {\it not}
incorporate any direct feedback (hydrodynamical or radiative) from the
``quasar-mode'' accretion (see Croton \etal 2006 for details).

The model uses the stellar population synthesis code of Bruzual \&
Charlot (2003) with the Chabrier IMF to predict observables such as
luminosities and colours in several filters, and also accounts for
dust extinction. However, the modelling of these additional
ingredients (especially dust) introduces further degrees of freedom
and uncertainties (see e.g. Fontanot \etal 2009b). For these reasons,
we prefer to limit the comparison between models and data to
metallicities and mass-weighted stellar ages. It should be noted,
however, that while these are direct outputs of the semi-analytic
model, these need to be estimated from observables in the case
of the data.

\begin{figure*}
\centerline{\psfig{figure=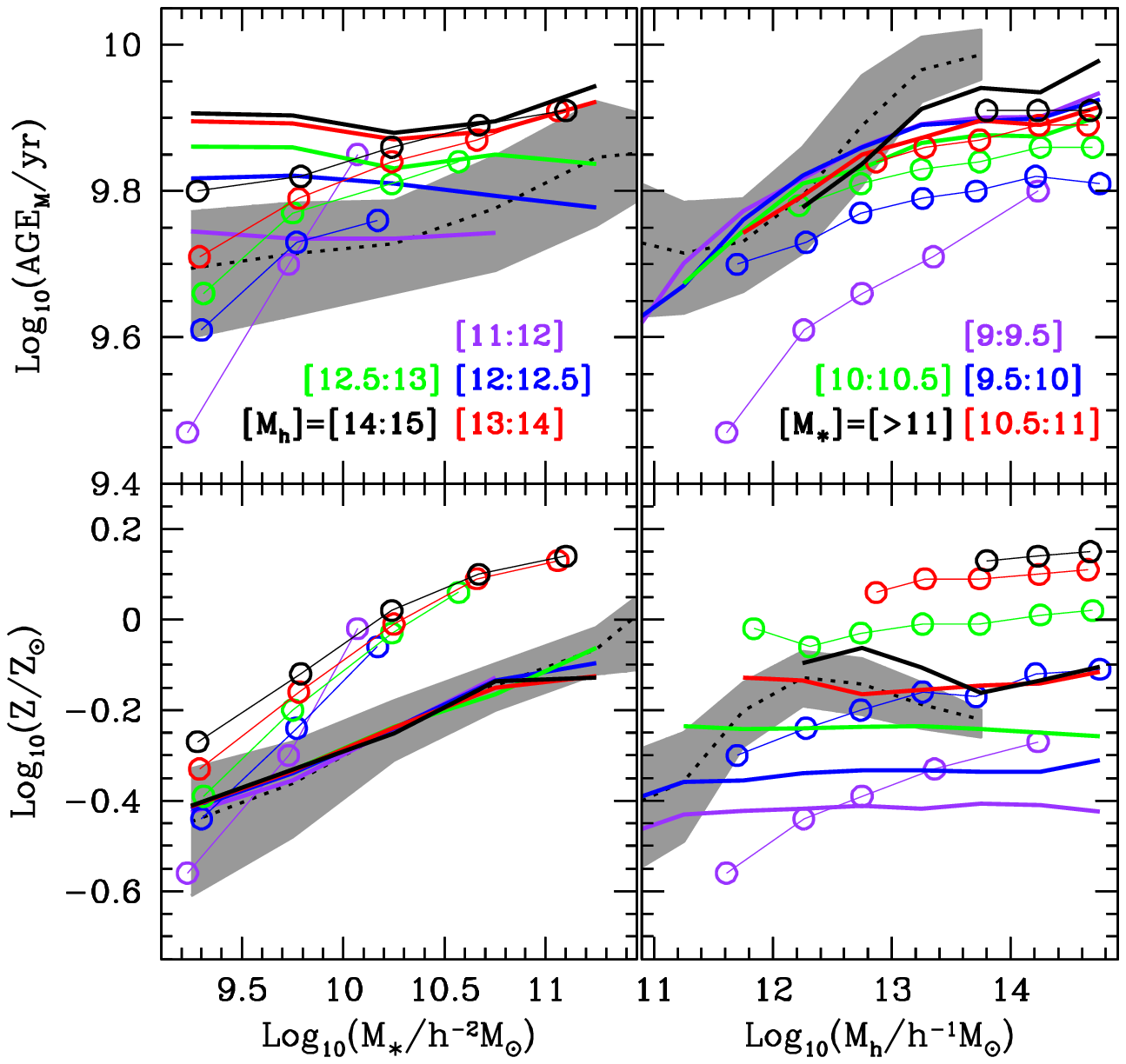,width=0.9\hdsize}}
\caption{The mass-weighted age and metallicity distributions of
  satellite and central galaxies as a function of stellar (halo) mass
  for different binnings in halo (stellar) mass from the SAM of Wang
  \etal (2008).  The colour-coding is the same as in
  Fig.\ref{fig:res}: solid lines trace the distributions of simulated
  satellites, while the grey band indicates the 16th-to-84th
  percentile range of centrals. Open circles correspond to the data
  shown in Fig.~\ref{fig:res} and are shown for comparison.
The typical scatter and error on the mean are as in Fig. 6.}
\label{fig:modelres}
\end{figure*}

\subsection{Model predictions}
\label{sec:modelpred}

The galaxy population simulated by W08 spans the same ranges in
stellar mass and halo mass as the observational sample analyzed here,
and consists of 62,075 centrals and 40,372 satellites. However, as
discussed in \S\ref{sec:sample}, our observational sample suffers from
a bias that originates from the fact that only galaxies whose spectrum
has a sufficiently high signal-to-noise can be used to determine their 
age and/or metallicity. As shown in Fig.~\ref{fig:sample}, this results
in a bias towards massive, red, early-type galaxies.  In order to
impose a similar bias onto the W08 results, we proceed as
follows. Using the complete sample II described in \S\ref{sec:sample}
we determine the fraction $f(S|M_{\ast},^{0.1}(g-r))$ of galaxies in
sample II that are in our $S$ sample, as function of both stellar mass
and $^{0.1}(g-r)$ colour. Next we accept each W08 model galaxy with a
probability equal to $f(S|M_{\ast},^{0.1}(g-r))$. Although this
results in removing a very significant fraction of the model galaxies
(our final sample consists of 11296 centrals and 7053 satellites), at
least the final sample of model galaxies is biased in a similar way as
the real data, thus allowing for a more meaningful
comparison\footnote{We have verified, though, that not applying this
bias correction yields results that are very similar, both
qualitatively and quantitatively.}. 

Fig.~\ref{fig:modelCS} shows the average metallicity and mass-weighted
stellar ages of the W08 model galaxies as functions of stellar and
halo mass, with central and satellite galaxies in grey and black,
respectively. The vertical bars in the left-hand panels indicate the
typical $\pm 1\sigma$ scatter in the model. For comparison, the grey
and black open circles indicate the observed values for centrals and
satellites in our sample $S$, respectively (cf. Fig.~\ref{fig:compCS}). 
For both observations and model, the typical error on the mean is 
smaller than a circle.
The upper left-hand panel shows that the
model overpredicts the stellar ages at the low mass end,
while underpredicting the ages of massive centrals. A particular
problem for the model seems to be the fact that it fails to reproduce
the steepening of the age-stellar mass relation at the low mass
end. We note that the mass-weighted age derived from a galaxy spectrum is 
still somewhat weighted by light, hence biased by the younger stellar
populations in the galaxy. This could partly alleviate the discrepancy
with the model. 
Nevertheless, the model yields satellites with stellar
populations that are older than those of centrals of the same stellar
mass. The age difference increases from $\sim 0.8$~Gyr at the massive
end to $\sim 2$~Gyr at $M_{\ast} = 3 \times 10^{9} h^{-2}\Msun$, in
good agreement with the data.  However, at intermediate masses
($M_{\ast} \sim 3 \times 10^{10} h^{-2} \Msun$), the model somewhat
overpredicts the age difference.

The upper right-hand panel of Fig.~\ref{fig:modelCS} shows that in
the W08 model central galaxies in massive halos ($M_{\rm h} \gta
10^{13} h^{-1}\Msun$) are $\sim 2$~Gyr older than their
satellites. Although in qualitative agreement with the data, the
central galaxies in massive halos are too old in the model compared
to the data (by $\sim 1.5$~Gyr).  In addition, the model predicts that
centrals and satellites in low mass halos ($M_{\rm h} \sim 10^{12}
h^{-1}\Msun$) are equally old, in disagreement with the data, which
shows that centrals are older than their satellites at all $M_{\rm h}$
probed\footnote{The lack
of centrals in bins of $M_{\rm h} \geq 10^{14} h^{-1}\Msun$ is
artificial, and due to the colour and stellar mass bias applied to the
model in Sect. 5.2.}.

Concerning stellar metallicities, the W08 model predicts that
satellite galaxies have, on average, the same metallicity as
centrals of the same stellar mass against
the observed trend that satellites become more metal-rich than
centrals at the low mass end. The model {\it does} reproduce the fact
that centrals are more metal rich than satellites for the same halo
mass, and by roughly the correct amount ($\sim 0.15$~dex), but the
model clearly underpredicts the stellar metallicities at the massive
end by about 0.2~dex, which is more than twice the $1\sigma$ scatter
in the model.

We now investigate what the W08 model predicts in terms of the stellar
ages and metallicities as functions of stellar mass at fixed halo
mass, and as functions of halo mass at fixed stellar
mass. Fig.~\ref{fig:modelres} is similar to Fig.~\ref{fig:res}, except
that it now shows the W08 model predictions, and only for the
mass-weighted ages and metallicities (as explained above, we do not
consider the model predictions for the luminosity weighted ages). The
dotted black line and the grey band correspond to the distributions
of model centrals (basically the same as in
Fig.~\ref{fig:modelCS}). The coloured thick lines represent the
average mass-weighted age (metallicity) of model satellites as a
function of stellar mass per bin of halo mass (left-hand plots), and
as a function of halo mass per bin of stellar mass (right-hand
panels).  For comparison, we have superimposed the data using open
circles (cf. Fig.~\ref{fig:res}, panels b, c, e, and f).

The comparison of model and data in Fig.~\ref{fig:modelres} reveals two
additional aspects in which the model does not reproduce the data. First of all, the
upper two panels show that the mass-weighted ages of model satellites
are virtually independent of stellar mass, but strongly dependent on
halo mass (with more massive halos hosting older satellites). For
massive satellites this is clearly in disagreement with the data,
which shows the opposite trend. For low mass satellites the
model and data agree in that the ages increase with halo mass,
although the model predicts a smaller age difference between
the lowest and highest $M_{\rm h}$-bins. Furthermore, the model 
overpredicts the ages of low mass satellites in low mass halos 
by $\sim 2.5$~Gyr.  


The second important shortcoming of the model is that it does not
reproduce the halo-mass dependence of the metallicities of low mass
satellites. The model predicts that satellite galaxies have
metallicities that depend on stellar mass (albeit with a slope that is
clearly too shallow), and are independent of the mass of the halo in
which they reside. The data, however, clearly show that the
metallicity of low mass satellites ($M_{\ast} \lta 3 \times 10^{10}
h^{-2} \Msun$) increases with increasing halo mass.
\begin{figure}
\centerline{\psfig{figure=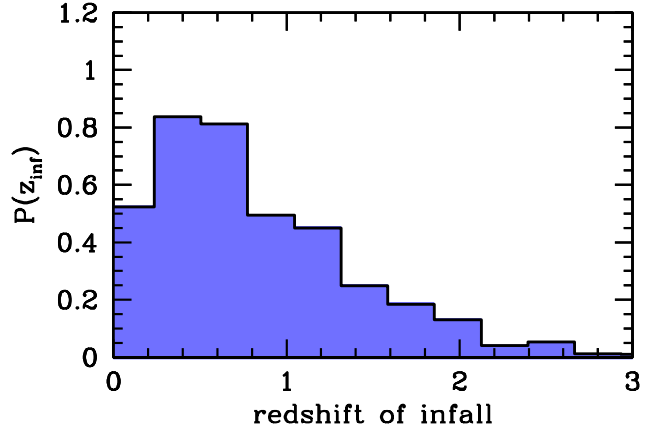,width=\hssize}}
\caption{The normalized distribution of the redshift of infall for
  present-day satellite galaxies in the W08 semi-analytical model. The
  redshift of infall is defined as the redshift at which the galaxy
  first became a satellite.}
\label{fig:zinfhis}
\end{figure}

To summarize, the semi-analytical model of W08 yields {\it relative}
ages and metallicities of centrals and satellites that roughly capture
the observed trends at the massive end.  However, the model does not 
reproduce several other aspects of the data presented here. In particular, 
the model
\begin{itemize}
\item[(i)] predicts an age-stellar mass relation that is much too shallow at
  the low mass end,
\item[(ii)] predicts that satellite galaxies in low mass halos have the
  same average, mass-weighted age as their centrals, in clear
  contradiction with the data,
\item[(iii)] yields metallicities at the massive end that are $\sim
  0.2$~dex too low, for both centrals and satellites,
\item[(iv)] predicts that centrals have the same metallicities as satellites
  of the same stellar mass, in disagreement with the data at the low
  mass end,
\item[(v)] predicts that satellites have mass-weighted ages that are
  independent of stellar mass, but strongly dependent on halo mass, in
  striking contrast to the data,
\item[(vi)] fails to reproduce the halo mass dependence of the metallicity
  of low mass satellites.
\end{itemize}

\subsection{The Evolutionary Paths of Centrals and  Satellites}
\label{sec:evolution}

In order to gain insight as to what might solve the problems listed
above, we first use the semi-analytical model of W08 to investigate
the differences in the evolutionary paths of centrals and satellites.
\begin{figure*}
\centerline{\psfig{figure=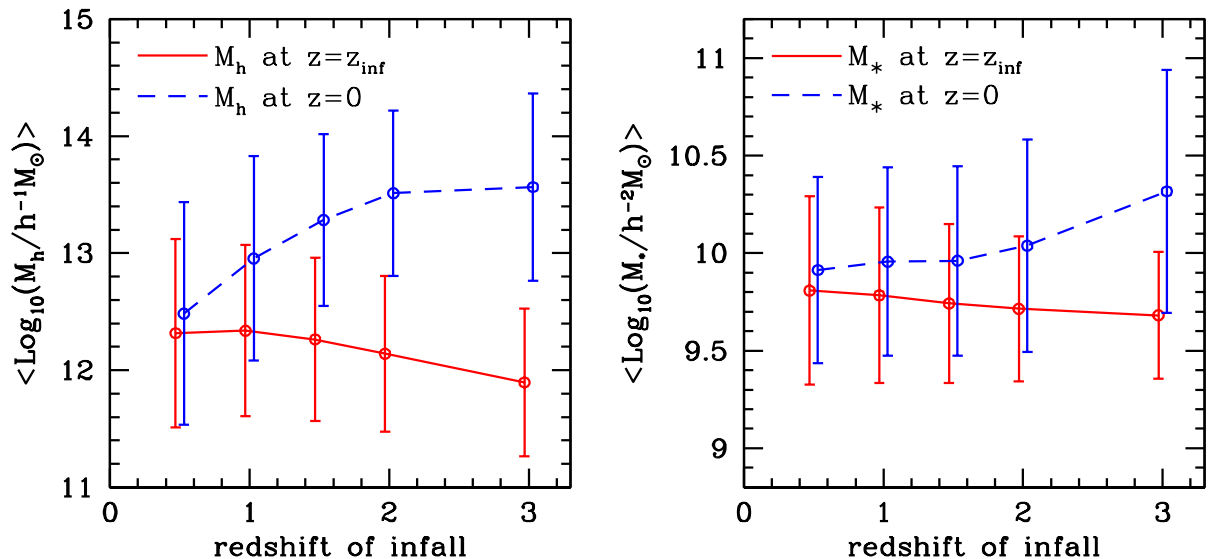,width=0.9\hdsize}}
\caption{{\it Left-hand panel:} the average halo mass of satellite
  galaxies in the W08 model at their time of infall (red, solid line)
  and at the present (blue, dashed line), as functions of the infall
  redshift. The errorbars reflect the scatter around the mean (the
  error on the mean is typically smaller than the symbol). For
  clarity, symbols have been slightly offset from their true redshift
  of infall. Note that the average halo mass at the time of infall
  only depends very weakly on $z_{\rm inf}$, while satellites with a
  higher redshift of infall at present reside in more massive halos.
  {\it Right-hand panel:} Same as left-hand panel, except that here we
  plot the average {\it stellar} mass of the satellite at time of
  infall (red, solid ilne) and at the present (blue, dashed
  line). Neither reveals a significant dependence on $z_{\rm inf}$.}
\label{fig:zinfall}
\end{figure*}

For this purpose we proceed as follows. For each model satellite
galaxy at $z=0$, we identify the redshift, $z_{\rm inf}$, at which it
first became a satellite. The distribution of these `redshifts of
infall' is shown in Fig.~\ref{fig:zinfhis}. Next we find a model
galaxy that remains a central at $z=0$, and that at $z_{\rm inf}$ has the
same stellar mass, metallicity and mass-weighted age as the satellite
at that redshift using the criteria: $\Delta$Log$_{10}(M_{\ast}) \leq$
0.1, $\Delta$Log$_{10}(AGE_M) \leq$ 0.1 and
$\Delta$Log$_{10}(Z/Z_{\odot}) \leq$ 0.2. These limits reflect the
typical uncertainty in estimating galaxy properties from SDSS data.

The solid line in the left-hand panel of Fig.~\ref{fig:zinfall} shows
the average halo mass of a satellite galaxy in this sample at its time
of infall as a function of $z_{\rm inf}$. There is a weak trend that
this halo mass increases with decreasing redshift, which mainly
reflects the evolution of the halo mass function (the average halo
mass increases with time). However, within the scatter, indicated by
the errorbars, this trend is extremely weak.  Thus, the typical mass
of the halo in which a satellite is accreted is virtually independent
of the time of accretion. However, the subsequent merger history of
their host halos is such that satellites with a higher $z_{\rm inf}$
are presently located in more massive halos, on average. This is
indicated by the dashed line, which shows the average satellite's halo
mass at the present ($z=0$). Clearly, satellites that were accreted
earlier (i.e. that became a satellite earlier) currently reside in
more massive halos, on average. The right-hand panel of
Fig.~\ref{fig:zinfall} is similar, except that it shows the average
stellar masses of the satellites, rather than their halo masses.
There is no significant dependence on $z_{\rm inf}$, neither for the
stellar mass of the satellite at infall, nor for its present-day
stellar mass. Note, though, that the average stellar masses at the
present are always somewhat higher than those at $z_{\rm inf}$,
indicating that satellite galaxies have continued to grow somewhat in
stellar mass since infall (at least in the W08 model considered here).

\begin{figure}
\centerline{\psfig{figure=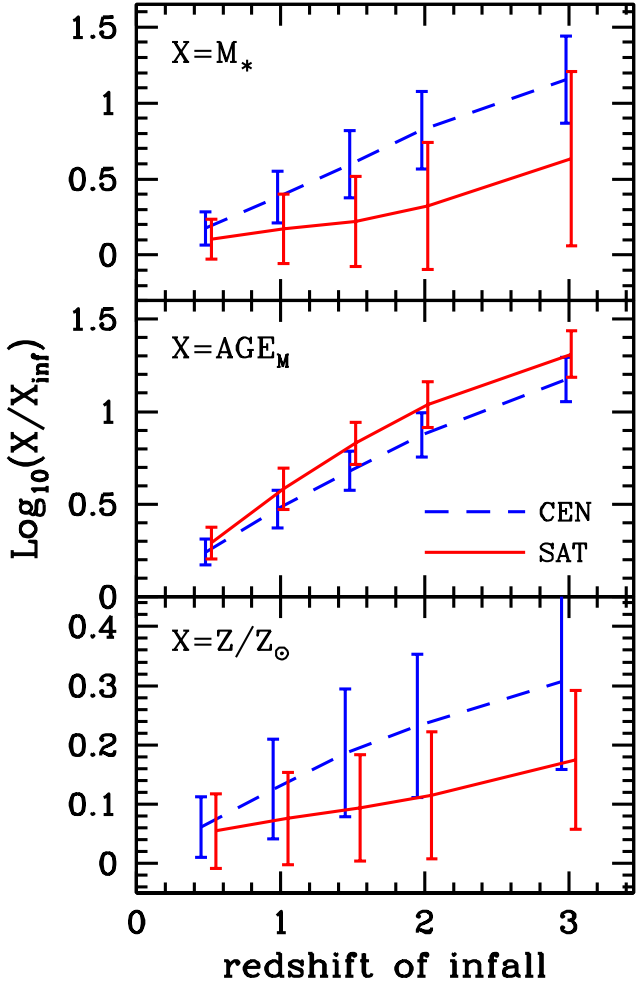,width=\hssize}}
\caption{The ratio of stellar mass (upper panel), mass-weighted age
  (middle panel) and metallicity (lower panel) at present to that at
  the time of infall of the satellite, as function of the redshift of
  infall. Results, obtained from the semi-analytical model of Wang
  \etal (2008), are shown for satellite galaxies (solid lines) and
  their paired centrals (dashed lines), where the pairing is such that
  the central and satellite have the same stellar mass, metallicity
  and mass-weighted age at the redshift of infall of the satellite.
  Lines and errorbars indicate the mean and $\pm$1$\sigma$ scatter,
  respectively.}
\label{fig:growth}
\end{figure}

Fig.~\ref{fig:growth} shows the difference in stellar mass (upper
panel), mass-weighted age (middle panel) and metallicity (lower panel)
between $z_{\rm inf}$ and the present day as a function of $z_{\rm
inf}$ for satellites (solid red lines) and centrals (dashed blue
lines). Errorbars reflect the $\pm 1\sigma$ scatter. As expected, the
growth in stellar mass, the increase in metallicity, and the aging of
stellar populations are all larger for higher $z_{\rm inf}$. More
importantly, this change is more pronounced for central galaxies than
for their paired satellites.  This is due to the way satellite
galaxies are treated in the semi-analytical model: once a galaxy
becomes a satellite, it is instantaneously stripped of its hot gas
reservoir.  Consequently, its star formation is rapidly quenched (how
rapid depends on its star formation rate, its cold gas reservoir, and
the strength of supernova feedback, which can expell the cold gas
before it has a chance to form stars). Without further star formation,
the stellar mass and metallicity can no longer increase, and the
stellar population ages passively. Galaxies that continue to be
centrals, continue to accrete new gas (unless some mechanism is
invoked to prevent this, such as AGN feedback). Hence, they continue
to grow in mass, continue to chemically enrich themselves (unless
feedback preferentially expells metals, as suggested, for example, by
the simulations of MacLow \& Ferrara 1999), and maintain relatively
young (luminosity-weighted) ages. The top panels of
Fig.~\ref{fig:growth} nicely confirm this picture; present-day
satellite galaxies have, to good approximation (and certainly within
the scatter), the same stellar mass and metallicity as they had at
their time of infall (at $z_{\rm inf}$). There is a trend that
satellite galaxies with a higher $z_{\rm inf}$ have managed to grow
more in stellar mass and metallicity than satellites with a lower
redshift of infall.  This mainly owes to the fact that satellite
galaxies continue to form stars until their cold gas reservoir is
exhausted, combined with the fact that galaxies at higher redshifts
have larger cold gas mass fractions. However, since the W08 model does
not include ram-pressure stripping, most likely real satellite galaxies
experience even less growth and enrichment after infall.  This,
combined with the fact that only a very small fraction of all
present-day satellites have $z_{\rm inf} > 2$ (see
Fig.~\ref{fig:zinfhis}), implies that it is reasonable to
postulate that satellite galaxies have not grown significantly in
stellar mass since their time of infall.

\begin{figure}
\centerline{\psfig{figure=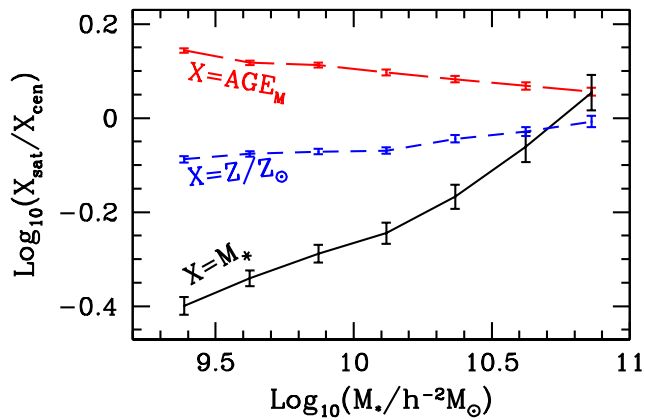,width=\hssize}}
\caption{The ratio in present-day stellar mass (black solid line),
  mass-weighted age (red, long-dashed line) and metallicity (blue,
  short-dashed line) between paired satellites and centrals in the SAM of W08
  as a function of the present-day stellar mass of the
  satellites. Errorbars indicate the error on the mean, not the
  scatter, which is substantially larger.}
\label{fig:samCS}
\end{figure}

Fig.~\ref{fig:samCS} plots the difference in {\it present-day} stellar
mass, mass-weighted age and metallicity between satellite galaxies and
their paired centrals as a function of the present day stellar mass of
the satellites.  Clearly, the differences between satellites and
paired centrals become larger for smaller $M_{\ast}$.  In particular,
present-day low mass satellites are less massive than their paired
centrals by as much as 0.4~dex, metal-poorer by $\sim 0.1$~dex, and
older by 0.15~dex.  At the massive end, however, satellites have
stellar masses, ages and metallicities that are very similar to those
of their paired centrals\footnote{Actually, satellites with $M_{\ast}
\gta 10^{11} h^{-2}\Msun$ appear to be more massive than their paired
centrals; this is due to the exponential tail of the stellar mass
function which causes a random central galaxy to have a stellar
mass that is almost always smaller than that of its paired satellite,
but still within 0.1 dex by construction.}.
Since there is little to no correlation between $z_{\rm inf}$ and
stellar mass (see Fig.~\ref{fig:zinfall}), this is not due to
satellite quenching. Rather, this trend is a consequence of AGN
feedback; massive centrals are quenched due to the action of AGN
feedback. The implementation of `radio-mode' AGN feedback is such that
it only becomes effective for central galaxies in sufficiently massive
halos. Hence, more massive centrals, which reside in more massive
halos, are quenched earlier. If a central galaxy that is paired to a
satellite is already quenched at $z_{\rm inf}$ (or will be quenched
soon thereafter), their stellar masses, metallicities and ages at the
present day will be similar. Low mass centrals have not (yet) been
quenched, so that they continue to form stars to the present day,
causing them to be younger, more massive and more metal-rich than
their paired satellites.

\section{Discussion}
\label{sec:disc}

The discussion in the previous section highlights that the relative
evolution of centrals and satellites in the semi-analytical model of
W08 is governed by two quenching mechanisms: AGN feedback, which only
operates on massive centrals, and strangulation, which operates, with
equal efficiency, on all satellites. Because of the way these
processes are treated in the model, the differences between centrals
and satellites depend on stellar mass, and on the redshift of
accretion, $z_{\rm inf}$.  Since present-day satellites in more
massive halos were, on average, accreted earlier, this also
introduces a dependence on halo mass.  Equipped with these insights,
we now discuss which modifications to the semi-analytical model may be
required in order to achieve better agreement with the data.

\subsection{Stellar Ages of Central Galaxies}
\label{sec:cenage}

We start by focussing on the (mass-weighted) stellar ages of central
galaxies. Here the W08 model reveals several problems: it predicts an
age-stellar mass relation that is too shallow, causing the model
centrals to have mass-weighted ages that are older at the low mass
end, and younger at the massive end. In addition, the model predicts
ages for centrals in massive halos that are older. At first sight,
since massive halos host massive centrals, these two problems seem to
be inconsistent with each other: how can massive centrals be 
younger, on average, but centrals in massive halos older?  The answer
to this paradox comes from the fact that the average relation between
stellar mass and halo mass of central galaxies becomes very shallow at
the high mass end (see e.g., Yang, Mo \& van den Bosch
2009a). Consequently, there is a large amount of scatter in halo mass
for a given stellar mass, but relatively little scatter in stellar
mass at a given halo mass (see More \etal 2009). In massive halos
($M_{\rm h} \gta 5 \times 10^{12} h^{-1} \Msun$) all centrals are
massive. They are also old, because of the implementation of AGN
feedback, which effectively shuts off any star formation in 
central galaxies in halos above a characteristic mass of $M_{\rm h}
\sim 5 \times 10^{12} h^{-1} \Msun$.  Because of the shallow slope of
the $M_\ast - M_{\rm h}$ relation, and because the halo mass function
is exponentially suppressed at the massive end, most massive centrals
with $M_\ast \sim 10^{11} h^{-2} \Msun$ still reside in halos in
which AGN feedback has not yet kicked in. Our results show that their
ages are younger compared to the data. 
This suggests that the problem with the ages of massive centrals
may be a reflection of the current implementation of AGN feedback not
being adequate. The data seem to
require a model in which the onset of AGN feedback has a more gentle
dependence on halo mass and/or stellar mass (see also Tinker \& Wetzel
2009).  Furthermore, the fact that the stellar ages of centrals in
massive halos are too old compared to the data, probably suggests
that AGN feedback is not as efficient as implemented in the W08 model.

The fact that the W08 model overpredicts the stellar
ages of low-mass centrals suggests that it yields star formation
efficiencies in low mass halos that are too high at high redshift 
(see also Fontanot et al. 2009a). A
similar conclusion is reached by Liu \etal (2009), based on a
comparison of different SAMs with data from the same SDSS galaxy group
catalogue as used here. The main mechanism in SAMs invoked to regulate
the star formation efficiency in low mass halos is supernova (SN)
feedback, suggesting that a modification of this feedback mechanism 
could be required. Alternatively, one can reduce the amount of star formation
in low mass halos by postulating that the high-redshift
intergalactic medium has somehow been preheated (e.g. Mo \& Mao 2002).

\subsection{Stellar Ages of Satellite Galaxies}
\label{sec:satage}

As is evident from Figs.~\ref{fig:modelCS} and~\ref{fig:modelres}, the
stellar ages of massive satellite galaxies in massive halos are in
good agreement with the data. However, the model overpredicts the ages
of low mass satellites and those of satellites in low mass
halos. Most likely this is a reflection of a related problem with
semi-analytical models, namely that they tend to overpredict the red
fraction of satellite galaxies (e.g. Weinmann \etal 2006b; Baldry
\etal 2006; Kimm \etal 2009). This has become known as the
`over-quenching' problem, and is believed to be due to an
oversimplified treatment of strangulation: whenever a galaxy becomes a
satellite galaxy, its reservoir of hot gas is assumed to be
instantaneously stripped and added to the hot component associated
with the central galaxy. Consequently, star formation in the satellite
is quenched shortly thereafter, once it has consumed (or expelled) its
remaining cold gas. However, detailed analytical and hydrodynamical
simulations have shown that the typical timescale for this stripping
process ranges from $\sim 1$ to 10 Gyr (e.g., Bekki, Couch \& Shioya
2002; McCarthy \etal 2008), and it has been shown that by descreasing
the efficiency with which hot gas is stripped from satellites, one can
obtain red satellite fractions in much better agreement with the data
(Kang \& van den Bosch 2008; Font \etal 2008; Weinmann \etal 2009).

Most likely, this will also solve the problem that the stellar
age-halo mass relation of the satellite galaxies in the W08 model is
shallow than for the data (upper right-hand panel of
Fig.~\ref{fig:modelCS}). This is due to the relation between infall
time and present day halo mass discussed in \S\ref{sec:evolution}. On
average, satellites in massive halos already became a satellite many
Gyrs ago. As long as the quenching timescale is small compared to the
time since infall, a modification thereof will have little impact on
the stellar ages. Satellites in low mass halos were accreted more
recently, on average, and their ages will therefore be more sensitive
to a modification of the quenching timescale.  Hence, increasing the
quenching timescale will steepen the stellar age-halo mass relation of
satellite galaxies, bringing it in better agreement with the data.

Note, though, that not all problems with the ages of satellite
galaxies are a reflection of the over-quenching problem.  In
particular, the fact that the model does not reproduce a stellar mass
dependence of the ages of satellite galaxies (upper left-hand panel of
Fig.~\ref{fig:modelCS}) is most likely a reflection of the fact that
the model does not reproduce the stellar mass dependence of the ages
of central galaxies. Since satellite galaxies were central galaxies
before they were accreted, it is clear that a correct reproduction of
the ages of satellite galaxies also requires the ages of central
galaxies to be correct.

\subsection{Metallicities of Central Galaxies}
\label{sec:cenmet}

The W08 model predicts a metallicity-stellar mass relation for central
galaxies that is clearly shallower than observed, resulting in
the metallicities of massive centrals being underestimated by $\sim
0.2$~dex. As shown in Bertone \etal (2007), the SAM of De Lucia \&
Blaizot (2007) suffers from exactly the same problem.  

One possibility is that this problem signals inadequate modelling of SN
feedback. Indeed, numerous studies have shown that the
mass-metallicity relation of (central) galaxies is highly dependent on
the exact treatment of SN feedback (e.g, Dekel \& Woo 2003;
Oppenheimer \& Dav\'e 2006; Bertone, De Lucia \& Thomas 2007; Brooks
\etal 2007; Finlator \& Dav\'e 2008; Dutton \& van den Bosch
2009). Possible modifications include, among others, a more dynamic
treatment of SN feedback (i.e., Gerritsen 1997; Monaco 2004; Bertone,
Stoehr \& White 2005; Stinson \etal 2006), or considering
momentum-driven winds rather than energy-driven winds (e.g., Murray,
Quatert \& Thompson 2005).  For example, Bertone \etal (2007) have
shown that incorporating the `dynamical' SN feedback model of Bertone
\etal (2005) in the SAM of De Lucia \& Blaizot (2007) yields a stellar
mass function and stellar metallicities in much better agreement with
the data. Furthermore, as already mentioned above, a modification of
the treatment of SN feedback may also be required in order to solve
the problem that the ages of low mass galaxies are too old.

It may also be possible to change the (stellar) metallicities of the
model galaxies by modifying the way metals returned to the
ISM via stellar winds and supernova are distributed over the cold gas
in the disk, the hot gas in the halo, and the galactic wind.  
In the W08 model all metals are simply deposited in the cold gas in the
disk (see e.g., De Lucia \etal 2004; Croton \etal 2006). However,
using hydrodynamical simulations, MacLow \& Ferrara (1999) have shown
that metals from the SN ejecta are far more easily ejected from a
galaxy than the gas (i.e., galactic outflows are predicted to be
strongly metal-enhanced with respect to the galaxy's ISM). This is
particularly true for low mass galaxies, i.e. $\sim$ 10$^9$ $M_{\odot}$,
where SN explosions can eject $\sim$70\% of the SN metals, thus 
preventing the cold gas and the next generation of stars from increasing
their metallicity by $\sim$0.2 dex. This ``missed'' metal enrichment
is consistent with the difference in stellar metallicity
between centrals and satellites at $M_{\ast} < 10^{10} h^{-2} M_{\odot}$
(see Figg. 4 and 5).

A somewhat more speculative alternative to modifying the prescription
for feedback is to consider an IMF that varies with redshift
(e.g. Larson 2005), with metallicity (e.g., Santoro \& Shull 2006), or
with the star formation rate (e.g., K\"oppen, Weidner \& Kroupa
2007). Making the IMF more top-heavy results in a higher yield, and
thus will effect the stellar metallicities.

Hence, there are a number of ways in which the SAM of W08 may be
modified so as to yield a metallicity-stellar mass relation for
central galaxies in better agreement with the data. 

\subsection{Metallicities of Satellite Galaxies}
\label{sec:satmet}

Of the six problems listed at the end of \S\ref{sec:modelpred}, three
involve the metallicities of satellite galaxies: problems (iii), (iv),
and (vi). Since satellite galaxies were central galaxies before they
were accreted, they basically inherit the metallicity problems for
centrals mentioned above. Hence, the fact that the W08 model severely
underpredicts the stellar metallicities of massive satellites [problem
(iii)], is simply a consequence of the fact that the model severely
underpredicts the stellar metallicities of massive centrals. However,
the other two problems are more satellite-specific: the failure of the
model to reproduce the fact that low mass satellites have higher
metallicities than central galaxies of the same stellar mass [problem
(iv)], and the failure to reproduce the halo mass dependence of the
metallicities of low mass satellites [problem (vi)].

The metallicity of a satellite galaxy is not expected to change
significantly once its star formation is quenched.  This is confirmed
by the lower panel of Fig.~\ref{fig:growth}, which shows that the
present day metallicity of a satellite galaxy is still very similar to
that at infall (at least for the vast majority of satellites with
$z_{\rm inf} \lta 2$).  This suggests two possible scenarios for
solving problems (iv) and (vi) which we now discuss in turn.

\subsubsection{Redshift evolution}

As we have seen in \S\ref{sec:evolution}, satellites in more massive
halos, on average were accreted earlier (i.e., became a satellite at
a higher redshift).  If indeed stellar metallicities evolve little
after being accreted, the observed halo mass dependence of the
metallicities of (low mass) satellite galaxies may simply be a
reflection of evolution in the metallicity-stellar mass relation of
(low mass) centrals.  Qualitatively, what is needed is a
metallicity-stellar mass relation whose zero-point {\it decreases} as
function of time (i.e., high redshift galaxies need to have a higher
metallicity than low redshift galaxies of the same stellar mass).
\begin{figure*}
\centerline{\psfig{figure=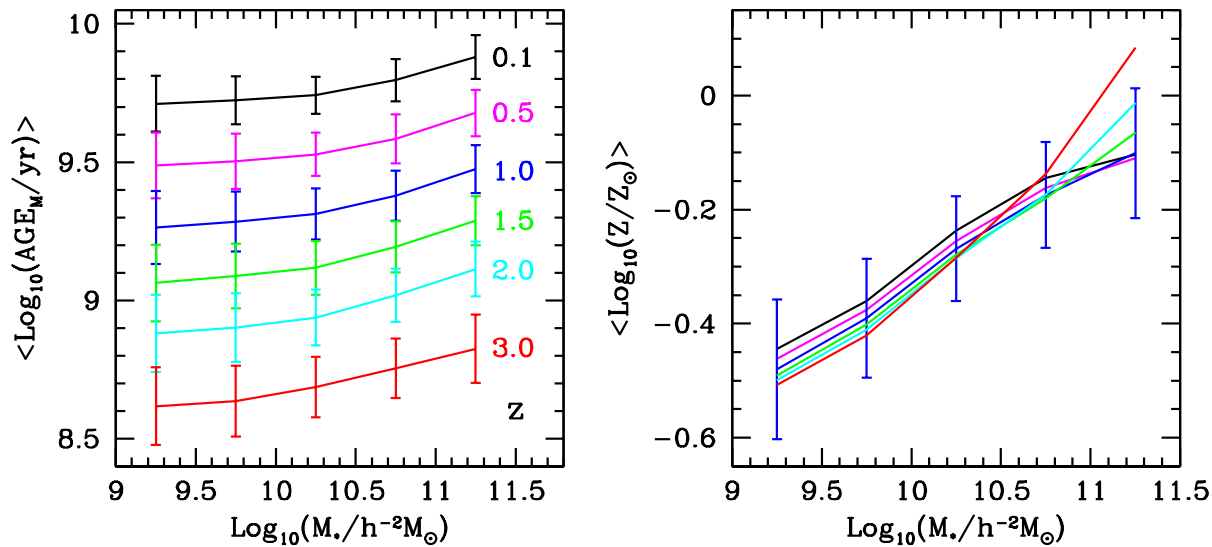,width=0.9\hdsize}}
\caption{Evolution in the age-stellar mass (left-hand panel) and
  metallicity stellar mass (right-hand panel) relations of central
  galaxies in the W08 model. Different colours correspond to different
  redshifts, as indicated in the left-hand panel.  Errorbars indicate
  the $1\sigma$ scatter in the relations.  For clarity, on the
  right-hand panel only the errorbars for the relation at $z=1$ are
  shown; those at other redshifts are very similar.}
\label{fig:evolution}
\end{figure*}

Observations of the gas-phase metallicity-stellar mass relation as
function of redshift, however, all seem to indicate that the
zero-point {\it increases} as function of time (e.g. Shapley \etal
2005; Savaglio \etal 2005; Erb \etal 2006; Maiolino \etal 2008).
Furthermore, using the spectra of SDSS galaxies, Panter \etal (2008)
find that also the {\it stellar} metallicities increase as function of
time (by more than a decade since $z \sim 3$ for galaxies with $M_\ast
< 10^{10} \Msun$). And finally, an increase in the zero-point of the
metallicity-mass relation with time is also supported by
hydrodynamical simulations of galaxy formation (e.g., Brooks \etal
2007).  Hence, we conclude that the observed halo mass dependence of the
metallicities of (low mass) satellite galaxies is not a consequence
of redshift-evolution in the metallicity-stellar mass relation.

Fig.~\ref{fig:evolution} shows how the age-stellar mass (left-hand
panel) and metallicity-stellar mass (right-hand panel) relations of
central galaxies evolve in the W08 model. The age-stellar mass
relation shows a very pronounced evolution of its zero-point (but not
of its slope), in that galaxies have younger stellar populations at
higher redshift, as expected. The metallicity-stellar mass relation,
however, shows only very little evolution, and this why the W08 model reveals 
no halo mass dependence in the metallicities of satellites.


\subsubsection{Tidal stripping}

A satellite galaxy and its dark matter subhalo are subjected to tidal
forces that cause mass loss. Initially, tidal stripping will only
remove the least-bound, outer regions of the dark matter subhalo.
However, as the satellite continues to lose orbital momentum due to
dynamical friction, and slowly sinks deeper and deeper into the host
halo's potential well, the tidal forces become stronger and may
ultimately cause the stellar component to experience mass loss as
well. Since more massive galaxies have higher metallicity, this may
cause satellite galaxies to end up having a metallicity that is `too
high' for their present day (remaining) stellar mass.  This would
explain why satellite galaxies have a higher metallicity than central
galaxies of the same stellar mass.

The data shows that the magnitude of this difference decreases as a
function of stellar mass (upper right-hand panel of
Fig.~\ref{fig:compCS}). One possible explanation could be that more
massive satellite galaxies experience less mass loss. However, we
consider this unlikely, given that more massive satellite galaxies
experience stronger dynamical friction, which causes them to sink
faster, and thus to be subjected to stronger tidal forces. A more
natural explanation is that the stellar mass dependence of the
metallicity difference $\Delta Z(M_\ast) \equiv Z_{\rm sat}(M_\ast) -
Z_{\rm cen}(M_\ast)$ is simply a consequence of the shape of the
metallicity-stellar mass relation of central galaxies. Since
\begin{equation}\label{dZdM}
{\Delta Z \over Z} \simeq \left({\rmd {\rm ln} Z\over\rmd {\rm ln}
  M_\ast}\right)_{\rm cen} \, {\Delta M_\ast \over M_\ast}\,,
\end{equation}
with $\Delta M_\ast$ the amount of stellar mass lost by the satellite,
it is clear that the corresponding metallicity difference depends on
the slope of the metallicity-stellar mass relation. Since the slope
$\rmd {\rm ln} Z/\rmd {\rm ln} M_\ast$ of the metallicity-stellar mass
relation of central galaxies becomes extremely small at the massive
end, even a large fractional mass loss results only in a tiny
difference in metallicity between central and satellite.  Furthermore,
since satellite galaxies in more massive halos were accreted (i.e.,
became satellites) at an earlier time (see \S\ref{sec:evolution}),
they have been subjected to tidal stripping induced mass loss for a
longer period. Consequently, satellite galaxies in more massive halos
should have a higher metallicity than satellite galaxies of the same
present-day mass in less massive halos, in qualitative agreement
with the results shown in the lower right-hand panel of
Fig.~\ref{fig:res}.

Hence, tidal stripping seems to give a fairly natural explanation for
both the stellar mass and the halo mass dependence of the metallicity
difference between centrals and satellites, at least
qualitatively. Quantitatively, we can use Eq.~(\ref{dZdM}) to estimate
how much stellar mass the average satellite galaxy must have lost in
order to explain the observed trend. The upper right-hand panel of
Fig.~\ref{fig:compCS} shows that satellite galaxies with a present-day
stellar mass of $M_\ast = 3 \times 10^9 h^{-2} \Msun$ have an average
metallicity that is $\sim 0.17$~dex higher than that of a central of
the same stellar mass. Using the slope of the metallicity-stellar mass
relation for centrals, this implies an average fractional mass loss of
50 percent!  Note, though, that this is only an underestimate. After
all, by using the present-day metallicity-stellar mass relation to
estimate the stellar mass of the satellite at infall (i.e., shortly
before it started experiencing mass loss), we are ignoring possible
redshift evolution. As discusssed above, observations suggest that
high redshift galaxies have lower metallicities than low redshift
galaxies of the same stellar mass. This implies that the metallicity
of a present-day satellite galaxy indicates an even larger stellar
mass at infall, and thus more stellar mass loss. It remains to be seen
whether such large fractional mass losses are realistic; it is
expected that once a (satellite) galaxy has lost a substantial
fraction of its original stellar mass due to tidal heating and
stripping, it becomes completely unbound (i.e., it is tidally
disrupted).  In fact, numerous studies in recent years have argued
that reconciling halo occupation statistics with halo merger rates
requires that a significant fraction of satellite galaxies is indeed
tidally disrupted (e.g., Conroy, Ho \& White 2007; Conroy, Wechsler \&
Kravtsov 2007; Kang \& van den Bosch 2008; Yang, Mo \& van den Bosch,
2009b).

As is the case for almost all semi-analytical models presented to
date, the W08 model does not include a prescription for tidal stripping
of satellite galaxies. Although it is tempting to identify this as the
reason for the model's failure to reproduce the observed halo mass
dependence of the metallicities of (low mass) satellite galaxies, it
remains to be seen whether a proper treatment of tidal stripping,
heating, and disruption, such as for example in Benson \etal (2002),
yields metallicities for satellite galaxies in better agreement with
observations. In particular, it may be challenging to reconcile the
relatively high disruption rates required to explain halo occupation
statistics with the number of surviving, but heavily stripped,
satellites required to fit the metallicity data presented here.

\section{Conclusions} 
\label{sec:concl}

We have combined the SDSS DR4 group catalogue of Yang \etal (2007)
with the catalogue of stellar ages and metallicities of SDSS galaxies of
Gallazzi \etal (2005) in order to study how the stellar ages and
metallicities of central and satellite galaxies depend on stellar mass
(the `nature' parameter) and halo mass (the `nurture' or `environment'
parameter).

Our findings can be summarized as follows:
\begin{itemize}

\item On average, satellite galaxies are older and metal-richer than
  central galaxies of the same stellar mass. This difference decreases
  with increasing stellar mass, becoming negligble for $M_\ast \gta
  10^{11} h^{-2} \Msun$.  At $M_\ast = 3\times 10^{9} h^{-2} \Msun$
  the average age and metallicity differences are $\sim 1.5$~Gyr and
  $\sim 0.15$~dex, respectively.

\item In absolute terms, the age differences between centrals and
  satellites are very similar when using luminosity-weighted or
  mass-weighted ages, indicating that they originate from differences
  in the integrated star formation histories.

\item On average, central galaxies are older (by $\sim 1$~Gyr) and
  metal-richer (by $\sim 0.15$~dex) than satellite galaxies residing
  in a halo of the same mass. Since central galaxies are more massive
  than satellite galaxies in the same halo, these differences simply
  reflect the relations between age/metallicity and stellar
  mass. Somewhat fortuitously, the average ages and metallicities of
  {\it all} galaxies (not distinguished between centrals and
  satellites) reveals no dependence on halo mass for $M_{\rm h} > 10^{12}
  h^{-1} \Msun$, in agreement with the studies of Sheth \etal (2006),
  Bernardi (2009) and Ellison \etal (2009).

\item A study of the ages and metallicities of satellite galaxies as
  functions of stellar mass at fixed halo mass, and as functions of
  halo mass at fixed stellar mass, reveals that the age and
  metallicity differences between centrals and satellites are largest
  for low mass satellites in massive environments. In particular, the
  average age and metallicity of low mass satellite galaxies ($M_\ast
  \lta 10^{10} h^{-2} \Msun$) increase with the mass of the halo in
  which they reside. For more massive satellites, both age and stellar
  metallicity are basically independent of environment (halo mass),
  but strongly dependent on stellar mass. Because of these effects,
  the slopes of the age-stellar mass and metallicity-stellar mass
  relations become shallower in denser environments (more massive
  halos).

\end{itemize}

In order to gain understanding of the physical origin of these trends,
we have compared our results with predictions of the semi-analytical
model (SAM) of Wang \etal (2008). This model predicts global galaxy
properties in good overall agreement with observations, both in the
local Universe and at higher redshift. In particular, it yields
stellar mass functions, two-point galaxy-galaxy correlation functions,
pairwise velocity dispersions, and a cosmic star formation history
that are all in very satisfactory agreement with the data.  However, a
comparison with the ages and metallicities presented here reveals a
number of important shortcomings.

Although the model predicts that satellite galaxies have older stellar
populations than central galaxies of the same stellar mass, in
qualitative agreement with the data, it yields metallicities that are
$\sim 0.2$~dex lower at the high mass end, and an age-stellar mass
relation that is too shallow. In addition, it fails to reproduce
the halo mass dependence of the metallicities of low mass satellites,
and predicts that centrals have the same metallicities as satellites
of the same stellar mass. We have also
compared the data presented here to the semi-analytical model MORGANA
(Fontanot 2009a,b), and found similar discrepancies between model and
data as for the W08 model. Given that the W08 model  
is very similar to those used in Kang \etal (2005), Kang, Jing
\& Silk (2006), Croton \etal (2006), De Lucia \etal (2006, 2007) and
De Lucia \& Blaizot (2007), we believe that the discrepancies found
in this paper are relatively generic for the concurrent generation of
semi-analytical models.

We have argued that the above mentioned discrepancies of the SAM indicate
the need to modify the implementations of both supernova feedback
(used to suppress and regulate star formation in low mass halos) and
AGN feedback (used to quench star formation of central galaxies in
massive halos). In addition, the models need to use more realistic
descriptions of strangulation (believed to be responsible for
quenching star formation in satellite galaxies), and a proper
treatment of the tidal stripping, heating and destruction of satellite
galaxies (which are completely ignored in almost all semi-analytical
models). Most likely, the combination of these improved recipes
will be able to  bring model predictions closer to observations.
The stellar ages and metallicities as function of stellar
mass and halo mass presented here may serve as a useful benchmark to
test and calibrate these ingredients. 

\section*{Acknowledgments}

We thank Simone Weimann and Andrea Macci\'o for useful discussions.
Some of the calculations were carried out on the PIA cluster of the
Max-Planck-Institut f\"ur Astronomie at the Rechenzentrum Garching.


\end{document}